\documentclass{aa}

\usepackage{graphicx}
\usepackage{epsfig}

\begin{document}

   \title{First image of a jet launching from a black hole accretion system: Kinematics}

   \author{B. Punsly}

   \institute{1415 Granvia Altamira, Palos Verdes Estates CA, USA
90274: ICRANet, Piazza della Repubblica 10 Pescara 65100, Italy and
ICRA, Physics Department, University La Sapienza, Roma,
Italy\\
\email{brian.punsly@cox.net}\\ }

   \date{Received March 12, 2023; }
\titlerunning{M\,87 Jet Kinematics}

\abstract{Jets are endemic to both Galactic solar mass and extragalactic supermassive black holes.  A recent 86 GHz image of M\,87 shows a jet emerging from the accretion ring around a black hole, providing the first direct observational constraint on the kinematics of the jet-launching region in any black hole jetted system. The very wide ($\sim280\mu\rm{as}$), highly collimated, limb-brightened cylindrical jet base is not predicted in current numerical simulations. The emission was shown to be consistent with that of a thick-walled cylindrical source that apparently feeds the flow that produces the bright limbs of the outer jet at an axial distance downstream of $0.4 \,\rm{mas}<z<0.65\, \rm{mas}$. The analysis here applies the conservation laws of energy, angular momentum, and magnetic flux to the combined system of the outer jet, the cylindrical jet, and the launch region. It also uses the brightness asymmetries of the jet and counterjet to constrain the Doppler factor. The only global solutions have a source that is located $<34 \mu\rm{as}$ from the event horizon. This includes the Event Horizon Telescope annulus of emission and the regions interior to this annulus. The axial jet begins as a magnetically dominated flow that spreads laterally from the launch radius ($<34 \mu\rm{as}$). It becomes super-magnetosonic before it reaches the base of the cylindrical jet. The flow is ostensibly redirected and collimated by a cylindrical nozzle formed in a thick accretion disk. The flow emerges from the nozzle as a mildly relativistic ($0.3c<v<0.4c$) jet with a significant protonic kinetic energy flux.}


 \maketitle
\section{Introduction}
The detection of a jet connected to the accretion flow in the radio galaxy M\,87 ($\approx 16.8$ Mpc distant) at 86 GHz with the Global Millimeter VLBI Array, the phased Atacama Large Millimeter/submillimeter Array, and the Greenland Telescope in 2018 provides the first direct evidence of a jet launching from a black hole accretion system \citep{lu23}. A limb-brightened cylindrical jet was found, $25 \mu\rm{as}< z< 100 \mu\rm{as}$, where $z$ is the axial displacement from the midline (see Figure 1) of the accretion ring \citep{lu23}. It is too wide and too collimated to be consistent with numerical simulations of Blandford-Znajek jets \citep{lu23}. The plasma source for the observed cylindrical jet was approximated by a thick-walled, uniform cylinder with a radius $R\approx 144 \mu\rm{as}\approx 38M$, and the tubular jet walls have a width $W \approx 36\mu\rm{as} \approx 9.5 M$ (where $M$ is the geometrized mass of the black hole), and it is apparently the source for the thick-walled tubular jet observed $0.405 \,\rm{mas}<z<0.65\, \rm{mas}$ downstream (see the left panel of Figure ). We call this the outer jet hereafter \citep{pun23,pun24}. The jet is detected within $z'\approx 20M$ above the midline of the accretion ring, where $z^{'}$ is the deprojected (intrinsic) distance. A line of sight (LOS) to the jet axis of $18^{\circ}$ and $3.8\mu\rm{as} =M\approx 9.5 \times10^{14}\rm{cm}$ is assumed throughout \citep{eht19}. This detection of a jet launching is a robust discovery that is of fundamental importance to the field of black hole jet-launching, regardless of future Event Horizon Telescope (EHT) imaging campaigns \citep{joh23}. Furthermore, if the cylindrical jet tapers to a launch point at smaller radii and the disk (beneath it) also radiates, then it might be impossible to unambiguously segregate the launch region from the disk proper due to the nearly face-on nature of the disk.  Hence, a kinematic analysis is likely necessary to achieve a deep understanding. This preliminary attempt is based on the currently existing data.
\par The main physical questions raised by this detection are: 1. why it was launched axially so far from the black hole, 2. how it attained such a high velocity, which was at least mildly relativistic, so close to the accretion plane when we assume bilateral symmetry and the lack of a counterjet detection, and 3. why the jet is so collimated. The motivation of this study is to answer these questions. The analysis proceeds as follows. Section 2 provides an estimate of the cylindrical jet flux density. Sections  3 and 4 explain the strategy of the analysis and the solutions, respectively. The last section discusses the results.

\begin{figure*}
\includegraphics[width= 0.42\textwidth]{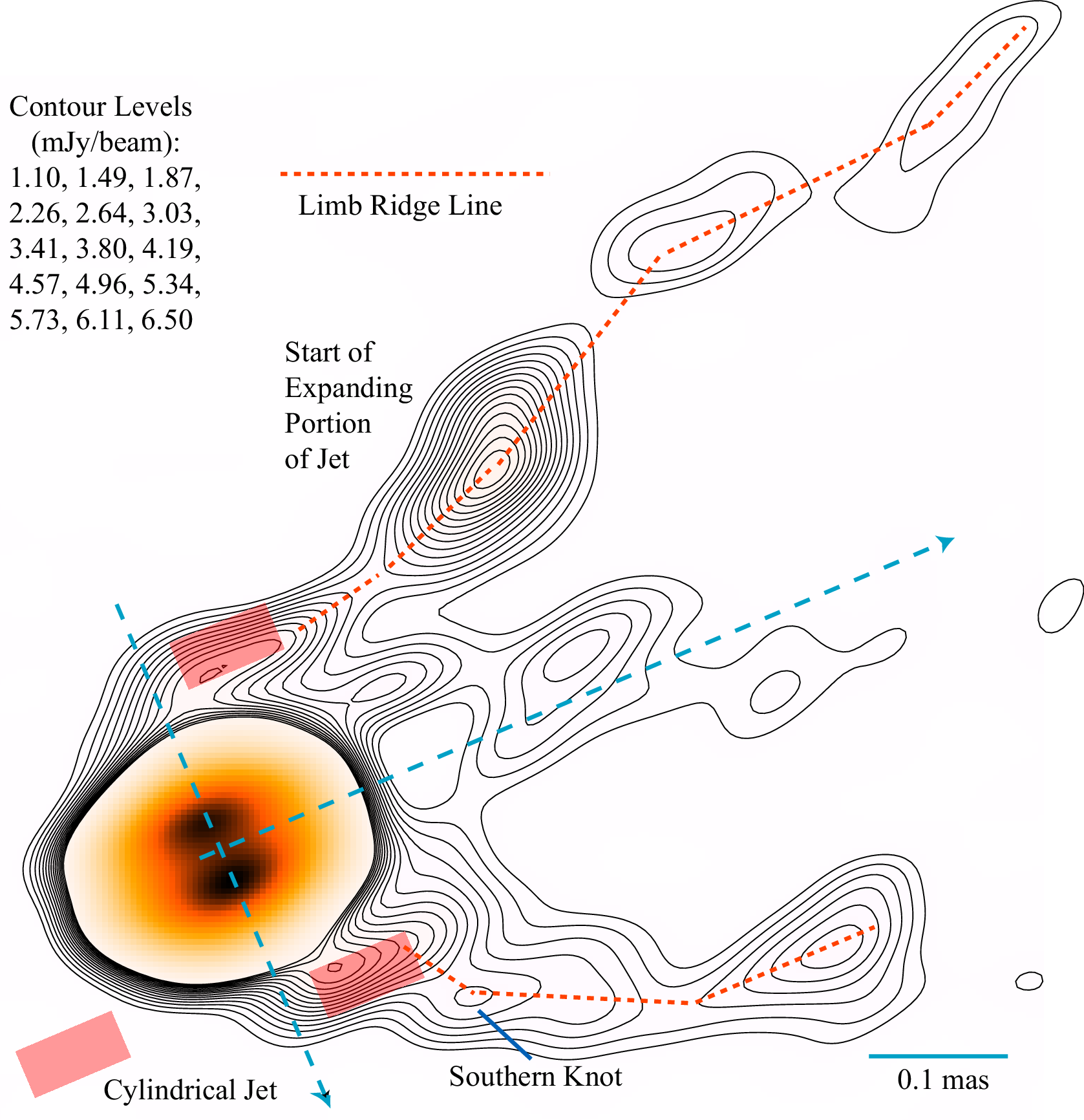}
\hspace{0.3cm}
\includegraphics[width= 0.5\textwidth]{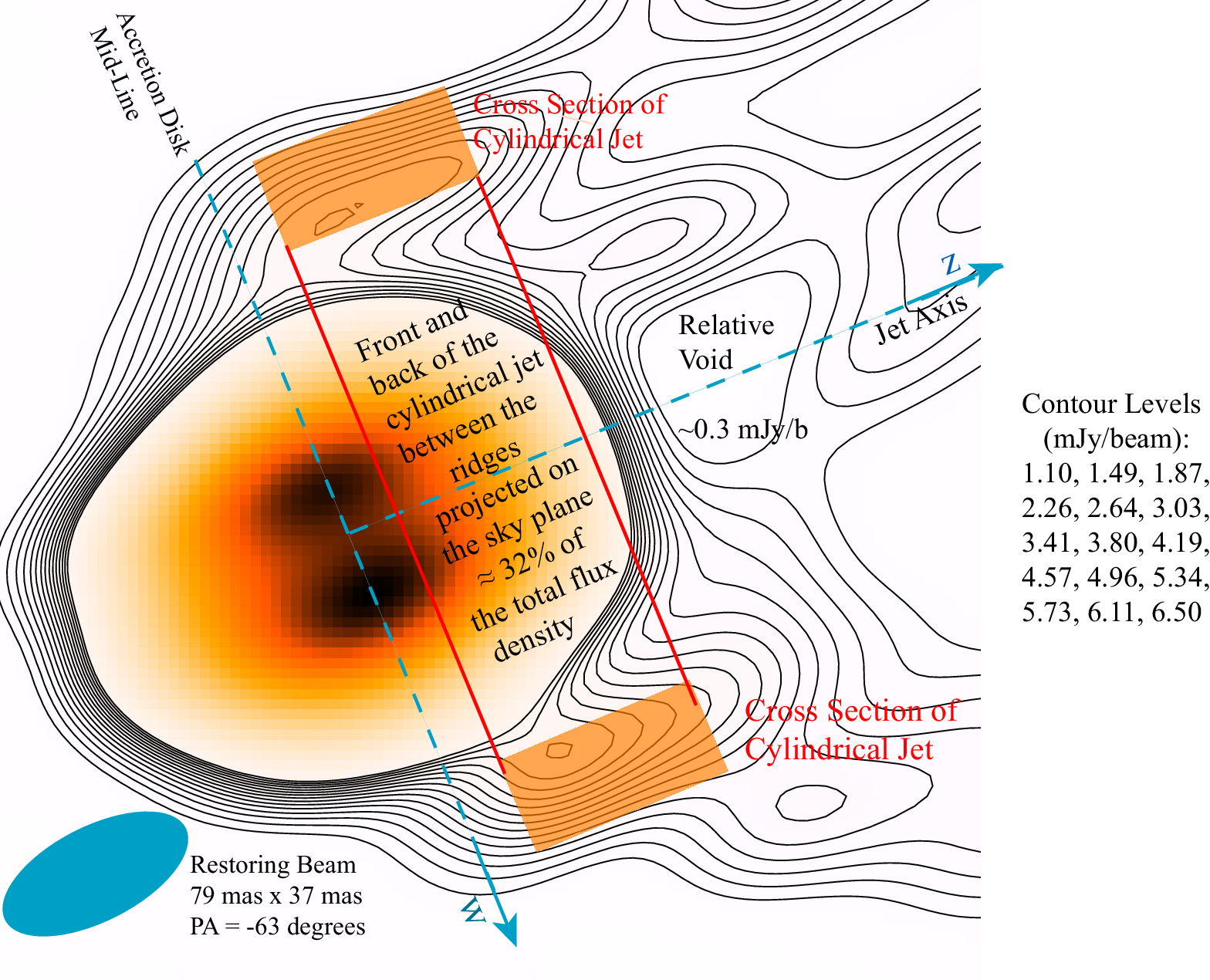}
\caption{Relating the cylindrical source to the cylindrical and outer jets. The left panel is a wide angle view from \citep{pun24}. This shows the cylindrical source as an extension of the ridges of the outer jet. The nucleus is saturated in this format, which accentuates the jet. The sky plane is the w-z plane and the midline of the putative accretion disk is the w-axis. The right panel is a close-up of the cylindrical source location. The location of the cylindrical source projected onto the sky plane is denoted by the red rectangle.}
\end{figure*}

\section{Flux density of the cylindrical jet base}
The right panel of Figure 1 is a reformatted version of the image FITS file used to create Figure 1a of \citet{lu23} that appeared in \citep{pun24}. The uniform cylinder source approximation that reproduced the cross sections of the cylindrical jet in \citet{pun24} has a flux density of $26.2^{+8.8}_{-7.1}$ mJy for $20\mu\rm{as}<z<100\mu\rm{as}$. No uncertainty was provided by \citet{lu23}. Thus, various crude, liberal uncertainty estimates were added in quadrature to create an overall liberal uncertainty estimate: the flux calibration (20\%), statistical noise (15\%), and the model (10\%) \citep{lee08,fom99}. The positive uncertainty includes the possibility that some emission (20\%) is resolved out by the interferometer beam. The average brightness ratio of the south to north ridge is $R(S/N)\sim 1.1$. However, the observations are not sensitive enough to determine small differences from unity like these due to the uncertainties associated with the statistical noise.

\section{Assumptions and solution strategy}The estimation of cylindrical jet properties and launch point is predicated on the assumptions of axisymmetry, bilaterally symmetry, perfect magnetohydrodynamics (MHD), and approximate time stationarity, combined with conservation of energy, $Q$, angular momentum, $L$, and magnetic flux, $\Phi$, with the outer jet. The cylindrical jet plasma was chosen to have uniform properties in the absence of a detection at higher resolution.

\par Approximate time stationarity is critical since the properties of the outer jet are required for the analysis. However, there are no observations at or near the epoch shown in Figure 1, April 14, 2018, with sufficient sensitivity for these purposes. The details of the outer jet in 2013 are reviewed in Appendix A. In order to justify the assumption of approximate time stationarity, we considered the nuclear light curves in Figure 2, whic h are an augmentation of the light curve in \citep{pun21}. The light curve combines the peak intensities from 43 GHz VLBA (with a 0.4 mas  x 0.2 mas restoring beam: $-16^{\circ}\leq \rm{PA} \leq 12^{\circ}$), fitted nuclear flux densities from 22 GHz VLBI Exploration Radio Astrometry (VERA) observations, and the 86 GHz Korean VLBI Network (KVN) nuclear flux densities, with an uncertainty $ < 20\%$ \citep{wal18,had12,had14,had16,kim18,alg21}. The 43 GHz VLBA observations have the highest resolution, but are not sampled densely enough, and the other measurements were designed to fill the gaps. The date of jet ejection is based on the average propagation velocity found in the next section. The light curve indicates that the M\,87 nucleus is typically steady within about $\pm 25\%$, and the nucleus might have been $\sim 20\%$ brighter when the 2018 outer jet was launched compared to the 2013 launch epoch \citep{kim18}. The nucleus receives a large contribution from the accretion disk, but theoretical models have a strong correlation between disk luminosity (accretion rate) and jet power \citep{pun08,aki22}. We concluded that using an observation in 2013 to estimate the outer jet $Q$ is probably not grossly inaccurate. To minimize this modest uncertainty, the estimates for the outer jet properties were biased toward the high side by choosing the $W/R\approx 0.35$ solution in Figure A.1. The assumption was not that $W/R=0.35$ was valid in 2018. The assumption was that the 2013 acceleration in the outer jet (see the upper right panel of Figure A.1) is qualitatively typical, and that the jet power is slightly stronger in 2018. Even though the exact jet properties are not known (i.e., the exact W/R, $Q$, and $L$), the choices are close enough for the results in Section 4 to be representative of M\,87 in 2018.

\par Viable solutions obey seven constraints on the global flow from the launch point, or anchor, for the magnetic field to the outer jet solution:
1. The flux density of the cylindrical jet, $S_{\nu}=26.2^{+8.8}_{-7.1}$ mJy; 2. $R(J/CJ)>5$ (see Equation 1); 3. $Q$ conservation; 4.$L$ conservation; 5. the jet requires an energy source. The cylindrical jet Poynting flux power of the cylindrical jet, $S^{P}$, exceeds that of the outer jet (magnetic energy converted into kinetic energy); 6. the mass flux does not decrease along the jet. Only entrainment is allowed during propagation, not the loss of mass flux; 7. $\Phi$ conservation.
\begin{figure}
\begin{center}
\includegraphics[width= 0.45\textwidth]{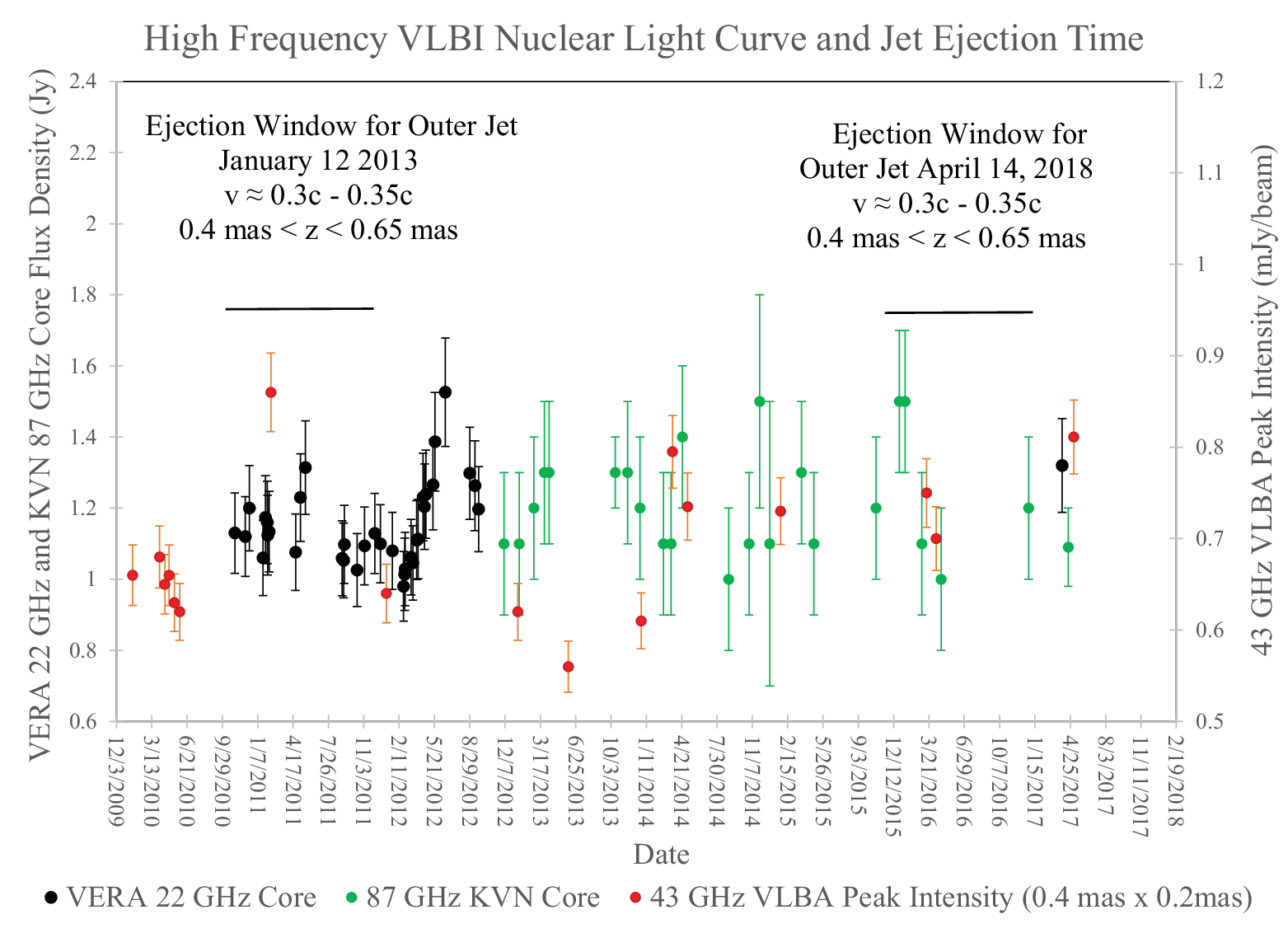}
\caption{Time variability of the nucleus. The figure shows the modest variability of the nucleus and motivates approximate time stationarity.}
\end{center}
\end{figure}
\begin{figure}
\includegraphics[width= 0.45\textwidth]{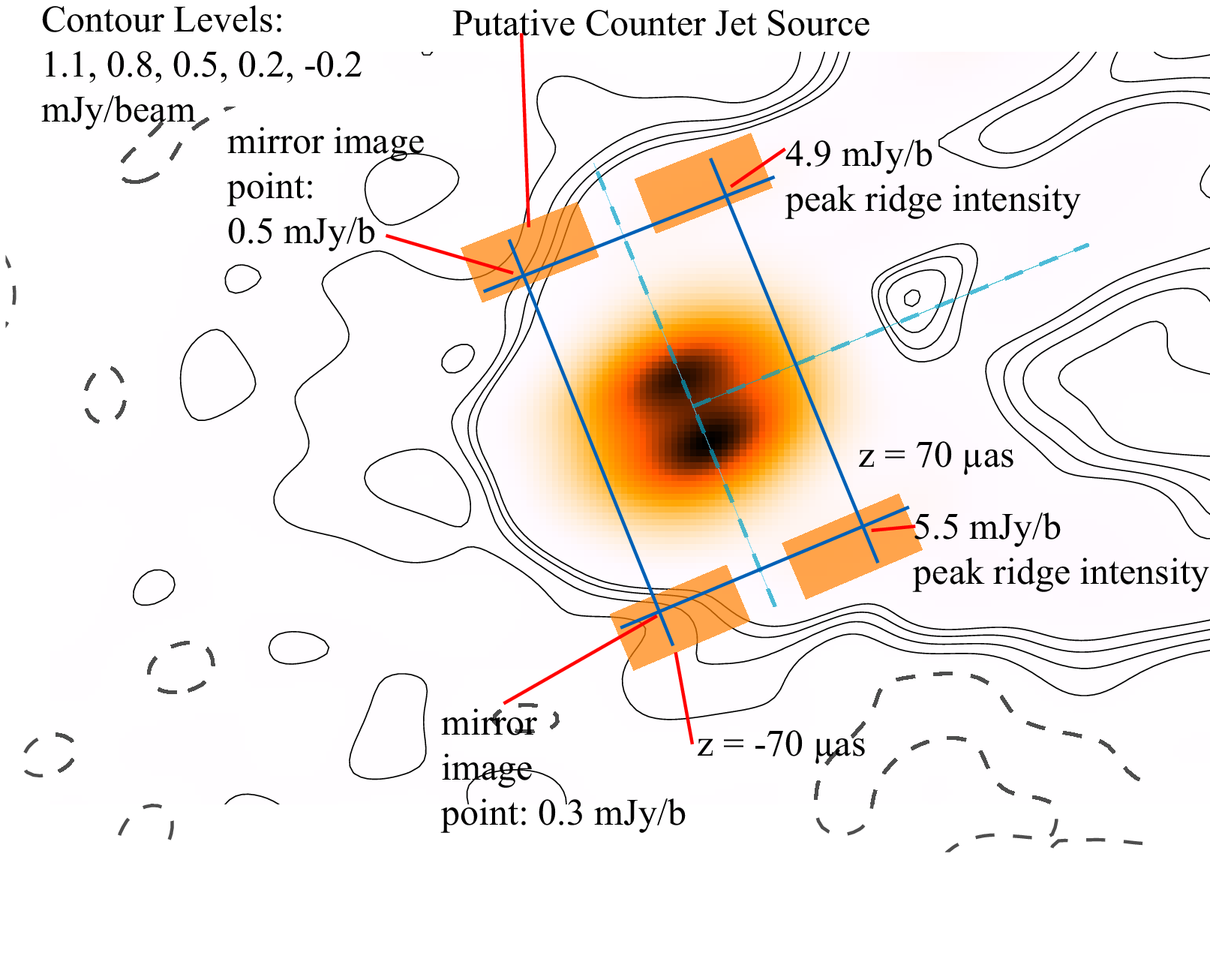}
\caption{Constraining the counterjet. The putative location of the counterjet cylindrical source. For $z<-70\mu\rm{as}$, the dilution from the bright nucleus is negligible at the ridge locations.}
\end{figure}
\section{Solution space}
Three different reference frames were needed for the calculations, the emission and observer's frame (subscripts e and o), and the cosmological rest frame of M87 (M87 frame hereafter). The calculations were performed in the emission frame and were then transformed to the observer's frame. The relevant equations are reviewed in Appendix B.
\par The upper limit on surface brightness of the counterjet ``within 0.1 - 0.3 mas" is $\sim1$ mJy/beam \citep{lu23}. Nuclear dilution affects our ability to resolve a weak counter jet. Figure 3 indicates that a 1 mJy/beam counter jet ridge would be cleanly resolved from the nucleus for $z <-70 \mu\rm{as}$ (0.14 mas from the black hole). Therefore, the  $z = 70 \mu\rm{as}$ peak intensities on the ridges that are indicated in Figure 2 motivate a lower bound of the brightness ratio, $R(J/CJ)> 5$. From Equation (B.3),
\begin{equation}
R(a/b) =\frac{[1+\vec{\beta_{a}}\cdot \mathbf{\hat{n}}]^{2+\alpha}}{[1+\vec{\beta_{b}}\cdot \mathbf{\hat{n}}]^{2+\alpha}}\;,
\end{equation}
where $R(a/b)$ is the flux density ratio, $\mathbf{\hat{n}}$ is the unit vector along the LOS, and $a$ and $b$ are two identical flows with different angles to the LOS. If $\vec{\beta_{a}} = -\vec{\beta_{b}}$ and $R(J/CJ)> 5$, the axial velocity of the cylindrical jet in the M87 frame is $\beta^{z'}>0.3$.

\par If the jet is launched from the disk, the field line angular velocity in the M\,87 frame, $\Omega_{F}$, would approximately be the local Keplerain angular velocity at which the poloidal field lines are anchored. In the M\,87 frame (Boyer-Lindquist coordinates, which are used throughout, unless otherwise stated), (t, r, $\theta$, $\phi$) this is \citep{lig75}
\begin{equation}
\Omega_{\rm{kep}}(r) = \frac{M^{0.5}}{r^{1.5}+aM^{0.5}}\;,
\end{equation}
where a is the angular momentum per unit mass of the black hole.

\par A method of graphically displaying conformance to conservation laws in the jet was developed in \citet{pun22} and repeated in Appendix A. These methods are applied to various potential global solutions of the cylindrical jet and the outer jet in Appendix C and summarized here. Figure C.1 shows conformance plots for the outer jet with various launch points. The only effect on the outer jet is the imposition of the $\Omega_{F}$ associated with the anchor. The top left example assumes that the jet emerges vertically from the local disk at $r=35$M, $\Omega_{F}=1.62\times 10^{-7} \rm{sec}^{-1}$ from Equation (2). Constraint (6) from Section 3 is grossly violated, and there is no global solution. The middle row shows improvement as the anchor point moves inward toward r =15 M. The outer jet fit is improved enough that it is now practical to expand the plot to include the cylindrical jet and $\Phi$ conservation. The result is nonconforming. $\dot{M}$ in the cylinder is below the plot, but still satisfies constraint (6). In the bottom left panel, anchor at r = 10M and a cylindrical jet velocity of v = 0.32c, the system conforms better. The standard of $\pm 5\%$ conformance is set by the outer jet in Figure A.1, it is not a pass or fail condition. The conformance plots are a diagnostic of the tension between conserved quantities induced by the assumptions of the solution. For v = 0.4c in the bottom panel, there is a significant $\Phi$ nonconformance. The fit deteriorates as the anchor point moves out because $\Omega_{F}$ decreases. From Equations (B.5), (B.7), and (B.9), this requires a higher azimuthal velocity,  $\beta^{\phi}$, to conserve $L$. A small $\Omega_{F}$ combined with a significant $\beta^{\phi}$ modifies the right-hand side of the last expression in Equation (B.4). The fact that $B^{P}/B^{\phi}\ll 1$ is inconsistent with the frozen-in condition in the outer jet induces large changes to the system of equations.
\par Figure C.2 shows the conformance plots as the footpoint (modeled as rings with constant plasma properties with $a/M=0.99$, as described in \citet{pun22}) moves inward. A model is required near the black hole because the gradients in $\Omega_{\rm{kep}}$ are large. $Q$ at the foorpoint is predominantly electromagnetic. Averaging over the ring, $\Omega_{F} = 7.04 \times 10^{-6} \rm{s}^{-1} = \Omega_{H}/2$ for r =1.7M -2.2M, where $\Omega_{H}$ is the angular velocity of the event horizon \citep{lig75}. Anchors with $r<10$M improve the $\Phi$ conformance for v=0.32c. For v=0.4c, $\Phi$ conservation is improved, but its conformance is just marginal. We conclude that v=0.4c is an upper bound on the velocity.
\par  The top left panel of Figure 4 considers Doppler asymmetry. $R(S/N)<1.2$ for launch points $<10\rm{M}$, consistent with the findings of Section 2. This is an independent corroboration of the conformance scatter plot analysis indicating $r < 10$M. The plasma angular momentum vector points antiparallel to the z-axis, the same as the EHT accretion ring \citep{eht19}. Note that based on the jet velocity and the finite size of the regions used to compute the flux density ($\approx$ beam full width at half maximum), the cylindrical (outer) jet data points in Figures C.1, C.2, and A.1 represent time averages with a duration of $\sim 3$ months ($\sim 6$ months) that smooth out abrupt fluctuations \citep{pun23}. The top right panel of Figure 4 shows the MHD wind properties in the cylindrical jet for different launch regions. The fastest MHD wave is the fast wave. In this analysis, the fast four-speed, $U_{F}^{2} \gtrsim (B^{p})^{2}/(4\pi m_{p}c^{2})$ \citep{pun08}. All the viable solutions (those launched inside of 10M) are superfast \citep{pun08}. Thus, the interaction of the magnetized flow with the disk at r =35M cannot be transmitted back to the launch region (it is causally disconnected). The next two rows plot the dependence on the physical properties of the cylindrical jet as a function of launch point and bulk velocity. In the bottom row, $B^{P}$ and $\beta^{\phi}$ are much larger with a launch point at r = 10M. This is a reflection of the tension among the MHD conservation laws, indicating that a viable solution breaks down at these larger radii.
\begin{figure*}
\begin{center}
\includegraphics[width= 0.45\textwidth]{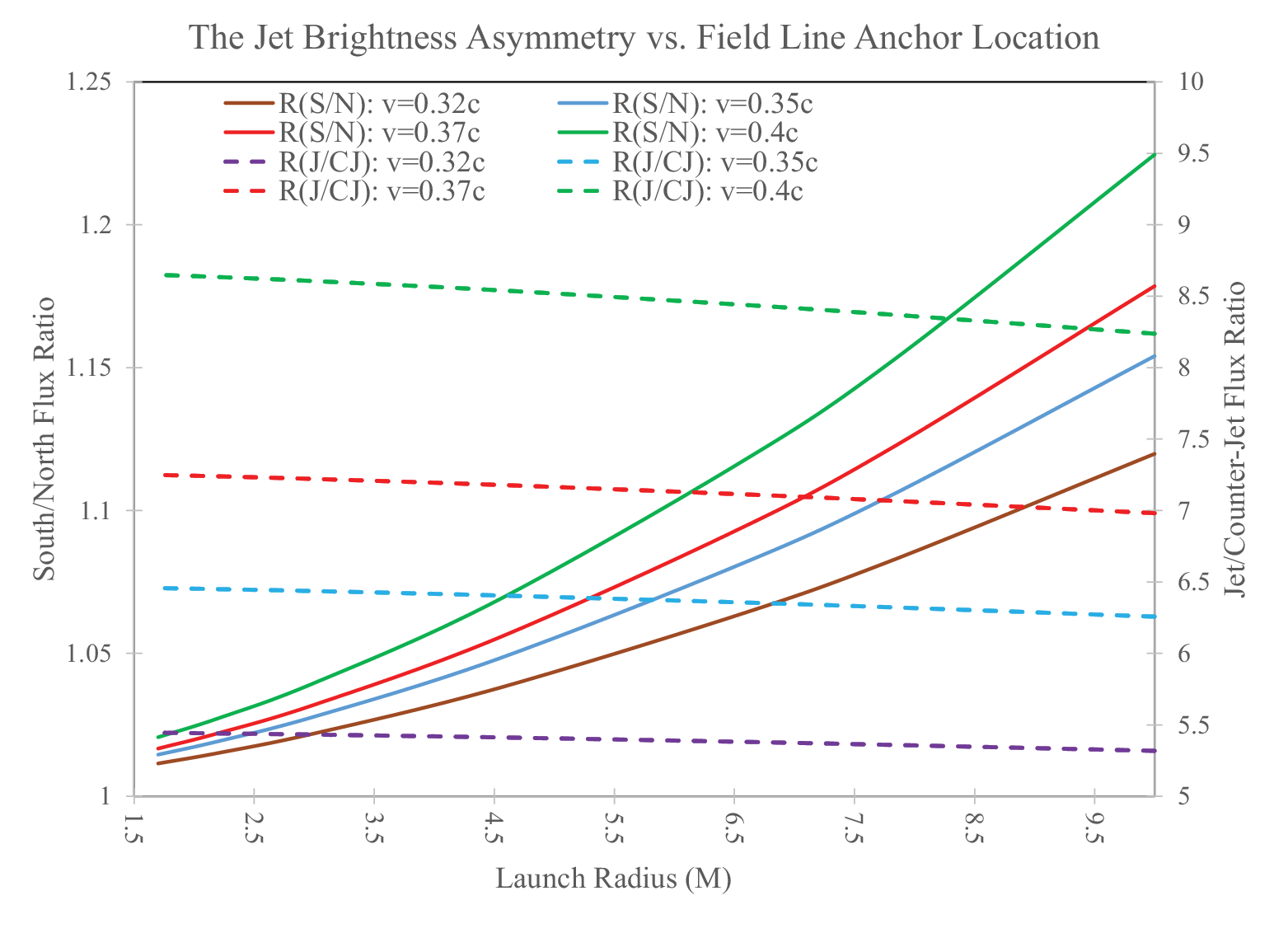}
\includegraphics[width= 0.45\textwidth]{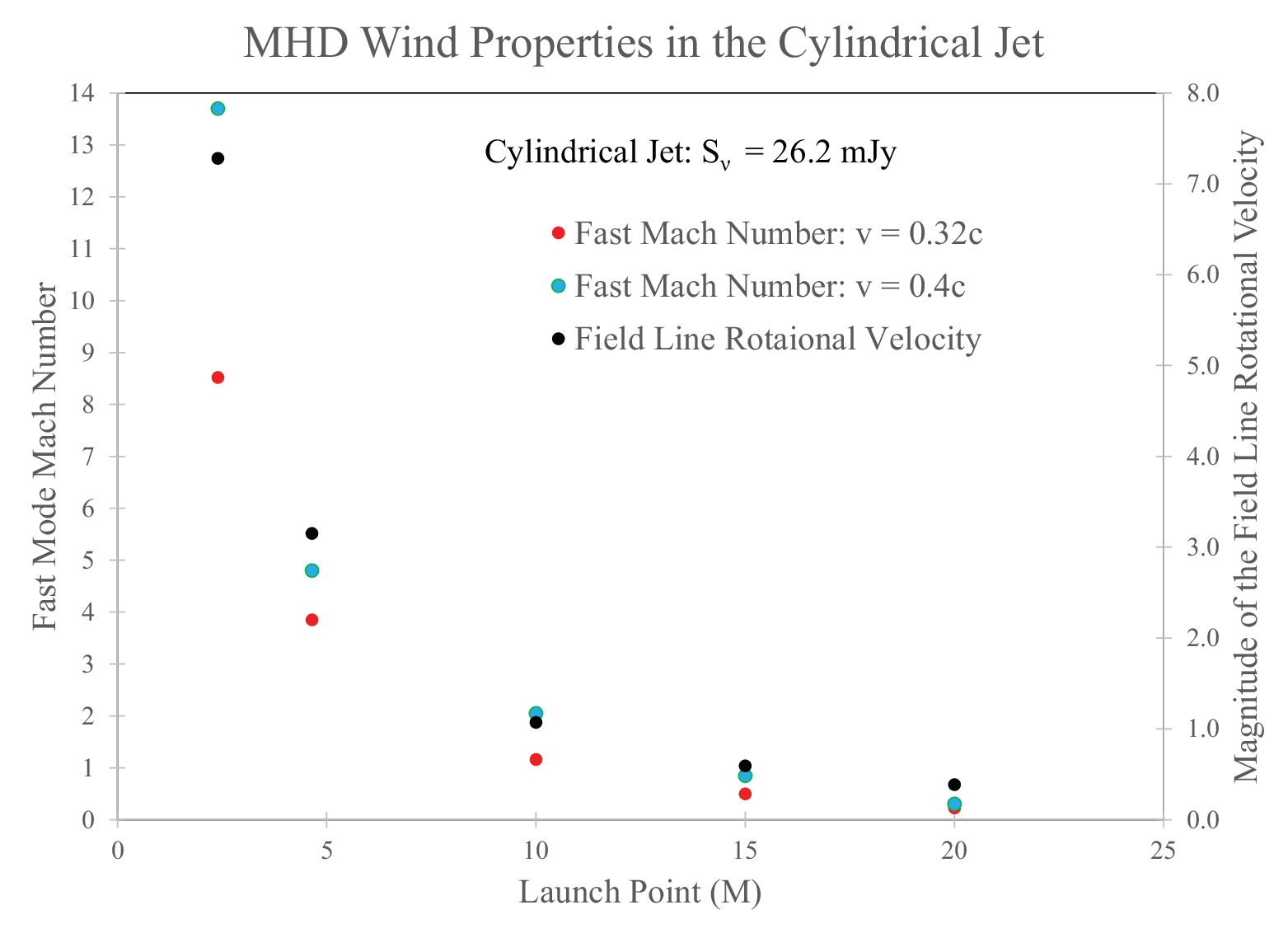}
\includegraphics[width= 0.45\textwidth]{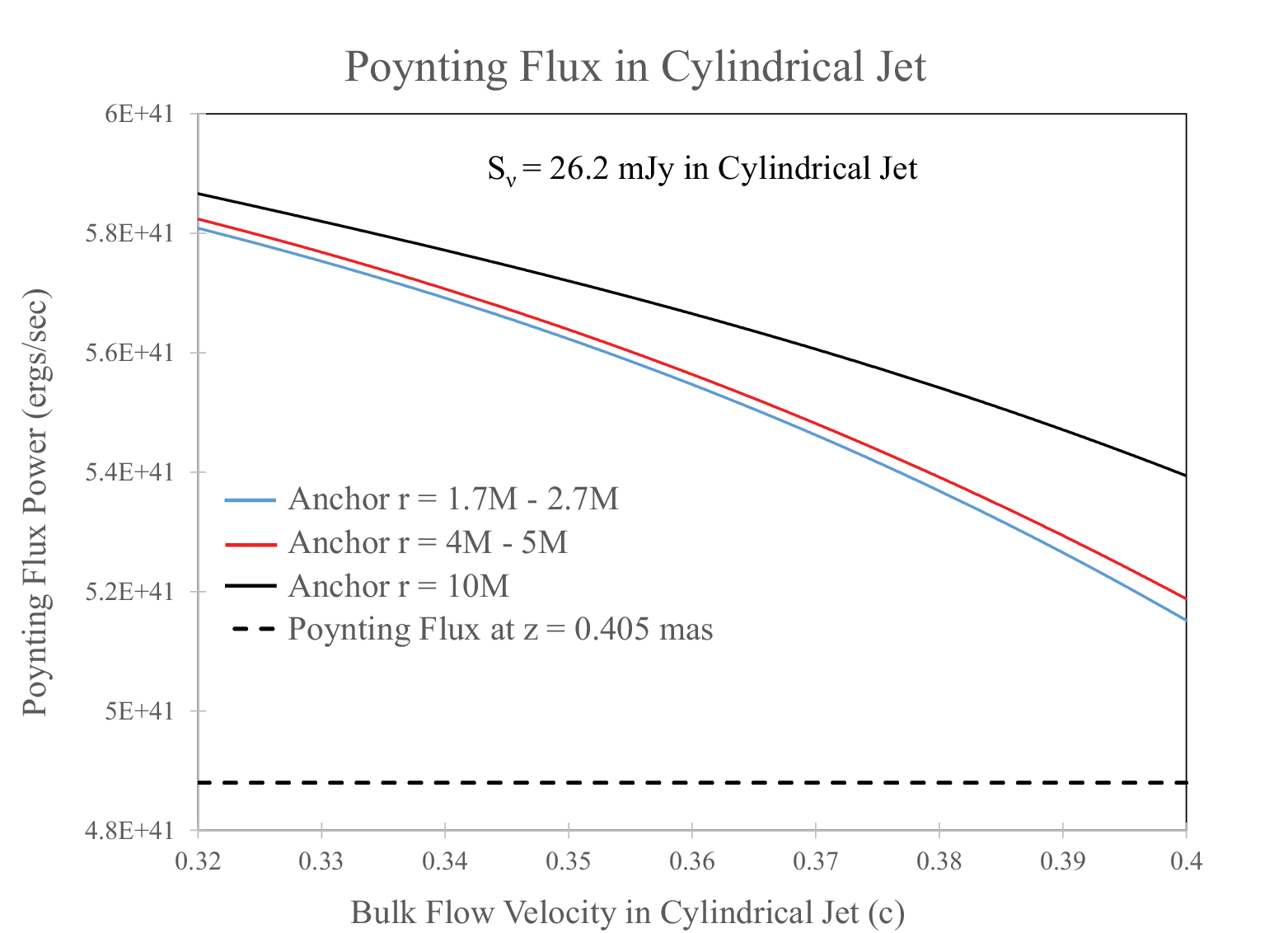}
\includegraphics[width= 0.45\textwidth]{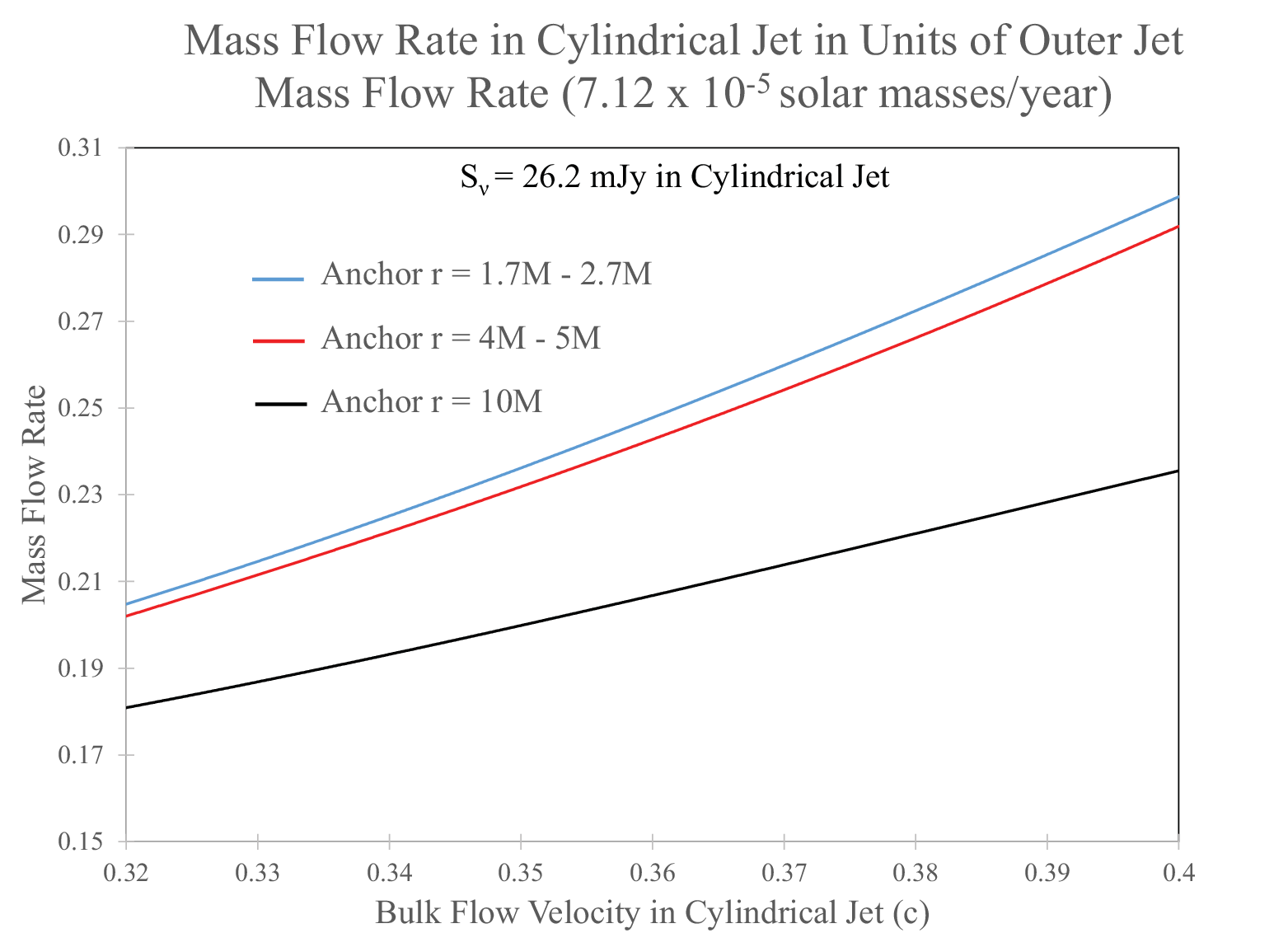}
\includegraphics[width= 0.45\textwidth]{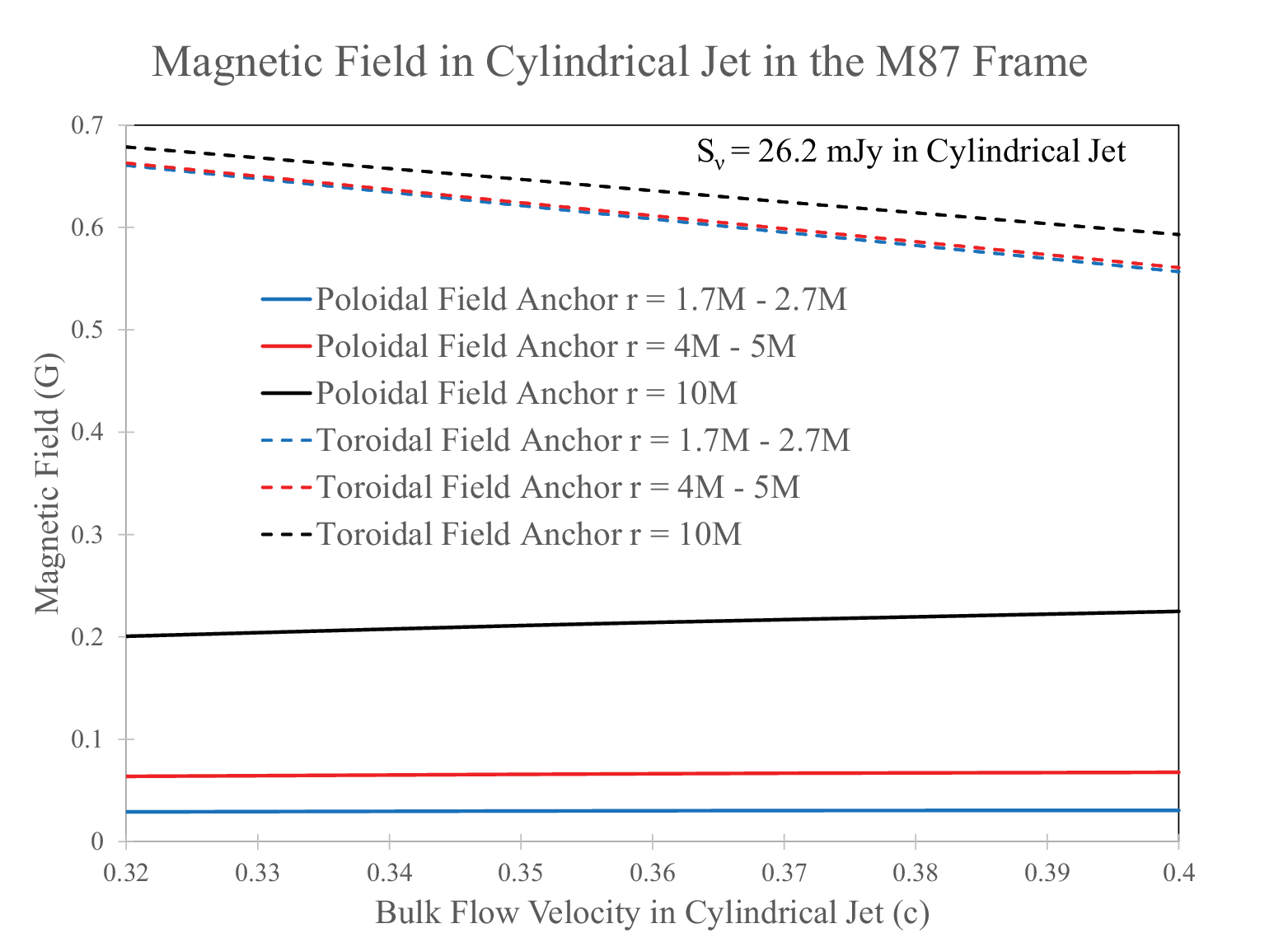}
\includegraphics[width= 0.45\textwidth]{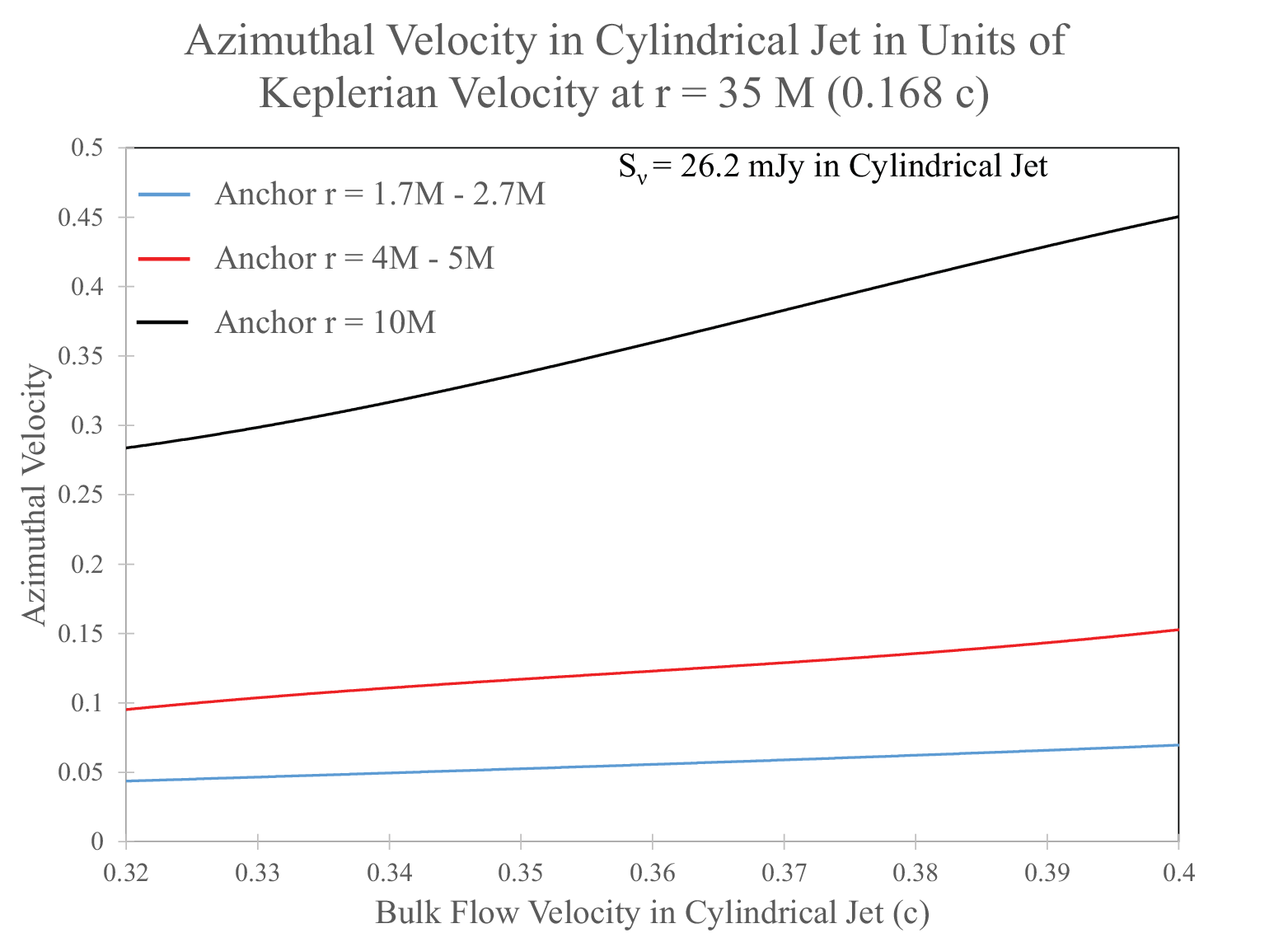}
\caption{Physical properties of the cylindrical jet. The top left panel indicates that the observed brightness asymmetries are consistent with the conforming solutions. The top right panel shows that all viable solutions are super-magnetosonic in the cylindrical jet.The physical properties of the cylindrical jet as a function of launch point and bulk velocity are given in the last two rows.}
\end{center}
\end{figure*}

\section{Discussion and conclusion}
This study explored the kinematics of the cylindrical jet base of M\,87 that was discovered in a 2018 observation. The analysis considered the global conservation of $Q$, $L$, and $\Phi$ in an attempt to answer the three questions posed in the Introduction. The global solutions indicate the following:
\par 1. The jet is launched axially so far from the black hole: The conformance plots in Appendix C show that i originates from small radii, $r<10$M, and expands laterally to r = 29.5M -39M. At this point, it transitions to an axial flow aligned with the outer jet. Apparently, plasma flows over the top of the disk and feeds the vertical jet. However, the inclination to the equator of the streamlines is likely never to be discerned due to the nearly face-on view of the accretion disk and lateral flow.
\par 2. It attained its high velocity so close to the accretion plane: The middle left panel of Figure 4 and Figure A.1 show that magnetic energy is converted into kinetic energy in the axial jet. When we extend this trend back toward the launch point,  plasma is accelerated by magnetic forces in the laterally expanding flow above the disk.
\par 3. The jet is so collimated: The strong collimation is much closer to the launch point than can be achieved in MHD jet models \citep{roh23,bla82}. It is almost axisymmetric, which also rules out the possibility of that it is a manifestation of sporadic nonaxisymmetric large-scale vertical flux ejection from reconnection at the interface of the disk with the event horizon magnetosphere \citep{rip22}. Furthermore, these flux tubes are not twisted, indicating that they transport negligible Poynting flux. By process of elimination, the straightforward explanation is that the axial jet appears to emerge from a nozzle in a thick accretion disk (half height to radius ratio, $h/r \approx 70\rm{M}/35\rm{M}= 2$). This seems to indicate a structured disk with different physical properties and $h/r$, an inner and outer disk.
\par The main uncertainty in this analysis is the flux density of the cylindrical jet. Appendix D shows that the results are unchanged for any flux density within the very liberal assignment of an uncertainty in Section 2. Another uncertainty is the outer jet properties. However, the solution is tightly constrained by observations with one unique solution for every $W/R$. Since $Q \sim (W/R)^{0.46}$, plausible $W/R$ values cannot generate a significant variation in $Q$ \citep{pun22}. These limitations mean that the analysis must rely on the light curve in Figure 2 to justify approximate time stationarity.

\par The mass flow rate is $\sim 25\%$ that of the outer jet in Figure 4. There must be entrainment of protonic plasma before it reaches the outer jet, probably by Kelvin-Helmholttz instabilities \citep{bic95}. The entrainment damps the electromagnetic acceleration of the plasma. The jet apparently strongly interacts with a non-negligible environment for the first $\sim 0.2$ light years of its propagation.
\begin{acknowledgements}
I am indebted to Thomas Krichbaum for explaining the uncertainties inherent to this remarkable image. I would like to thank Rusen Lu for the image FITS file. This manuscript benefitted from the improvements suggested by the supportive referee.
\end{acknowledgements}

\begin{appendix}
\section{Review of the details of the outer jet}
The only observations for which a reliable estimate of the outer jet kinematics can be obtained are the high-sensitivity 43 GHz VLBA observations from January 12, 2013 \citep{wal18}. This image alone clearly defines the limb-brightened counterjet (top left panel in Figure A.1). A high-sensitivity depiction of the counterjet is essential for the kinematic analysis. Doppler-beaming calculations (assuming intrinsic bilateral symmetry) in \citet{pun21} indicate consistency with a uniformly accelerating jet from 0.27c to 0.38c in the range $0.4\, \rm{mas}<z< 0.65 \,\rm{mas}$ (top right panel in Figure A.1). This bulk velocity is more relevant for defining the large-scale kinematics than anecdotal pattern speeds (e.g., MHD wave speeds or local inhomogeneities).
\par An analysis of the outer jet based on the bulk velocity field, flux density distribution, and the conservation laws of mass, energy, and angular momentum appeared in \citep{pun22}. Various cases were considered. The first case was a minimum-energy plasma in which the magnetic field was turbulent and the plasma protonic. The jet power, $Q$, increased significantly and the mass flow rate, $\dot{\mathcal{M}}$, decreased significantly along the jet. Thus, this assumption did not satisfy energy or mass conservation and was disfavored. The same results were found for a positronic plasma instead of a protonic plasma. The outer jet plasma is not in a minimum-energy state.
\par Next, a nearly pure Poynting flux outer jet was assumed. Even though $Q$ was conserved by assumption, $\dot{\mathcal{M}}$ decreased significantly along the jet. Thus, this assumption did not satisfy mass conservation. The result holds for a positronic or a protonic plasma. The outer jet is not a Poynting jet.
\par The only mechanism that can ensure mass conservation (or any a nondecreasing mass flux as the jet propagates, for that matter) with the other constraints satisfied is a significant conversion of magnetic into mechanical energy in which the kinetic energy flux is non-negligible in comparison with the Poynting flux. This solution requires a protonic plasma. The approaching jet kinematic properties are shown in the middle panels of Figure A.1, originally from \citep{pun22}. In summary, the tubular jet was determined to be a mildly relativistic protonic jet.
\par The solution is completely constrained (kinematics and composition) by the observation for each value of $W/R$. The tubular jet wall was estimated at $W\approx 0.25R$, same as the cylindrical jet \citep{pun24}. It was pointed out by \citep{wal18} that the LOSs that graze the inner wall of the tubular jet (the longest LOSs through the jet wall) should be near the peak intensity of the ridges. This was associated with the inner wall of the tubular jet. The width, $W$, of the tubular jet wall was crudely estimated by fitting a Gaussian line source, after convolution with the restoring beam, to the cross-sections of the approaching jet in the image in the top left panel of Figure A.1. The half width at half maximum of the Gaussian source was assumed to be a decent approximation to $W$.

\par Based on the above, an accurate estimate of $W/R$ would improve the estimate of the outer jet properties. Thus motivated, numerous axisymmeric jet models were investigated in a detailed analysis of high-sensitivity high-resolution jet images in 2013 and 2014 \citep{pun23}. It was found that $W/R=0.35$ in the axisymmetric models provided the best fit to the cross-sections of the images \citep{pun23}. The estimates in the original solution for $W/R$ range from 0.24-0.27. They are all increased by a factor of 1.4 =0.35/0.25, and the solution is recalculated in the bottom panels of Figure A.1. The two solutions behave almost identically. The conformance plots on the right are the main tool we used to assess the fit quality in the body of the manuscript.
\begin{figure*}
\begin{center}
\includegraphics[width= 0.45\textwidth]{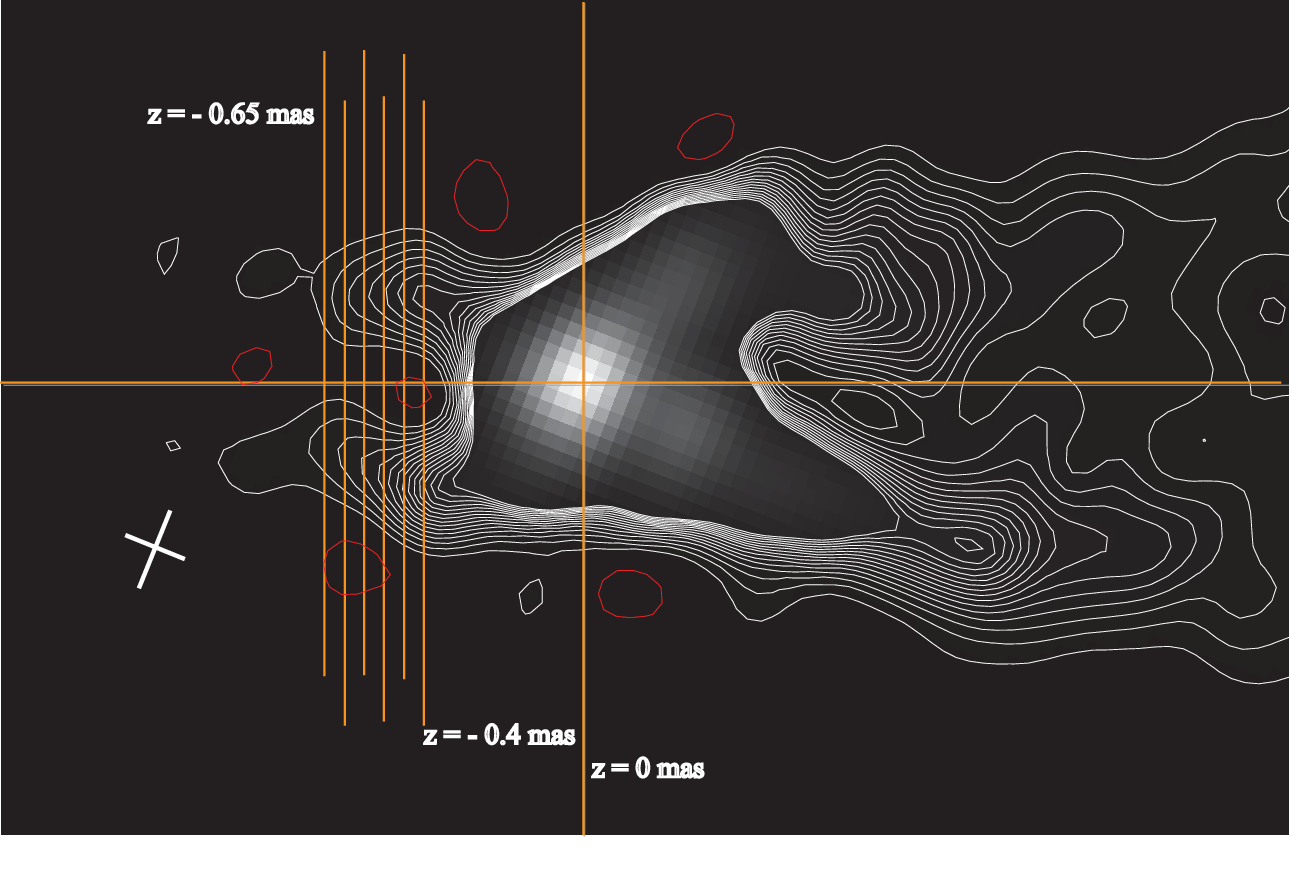}
\includegraphics[width= 0.45\textwidth]{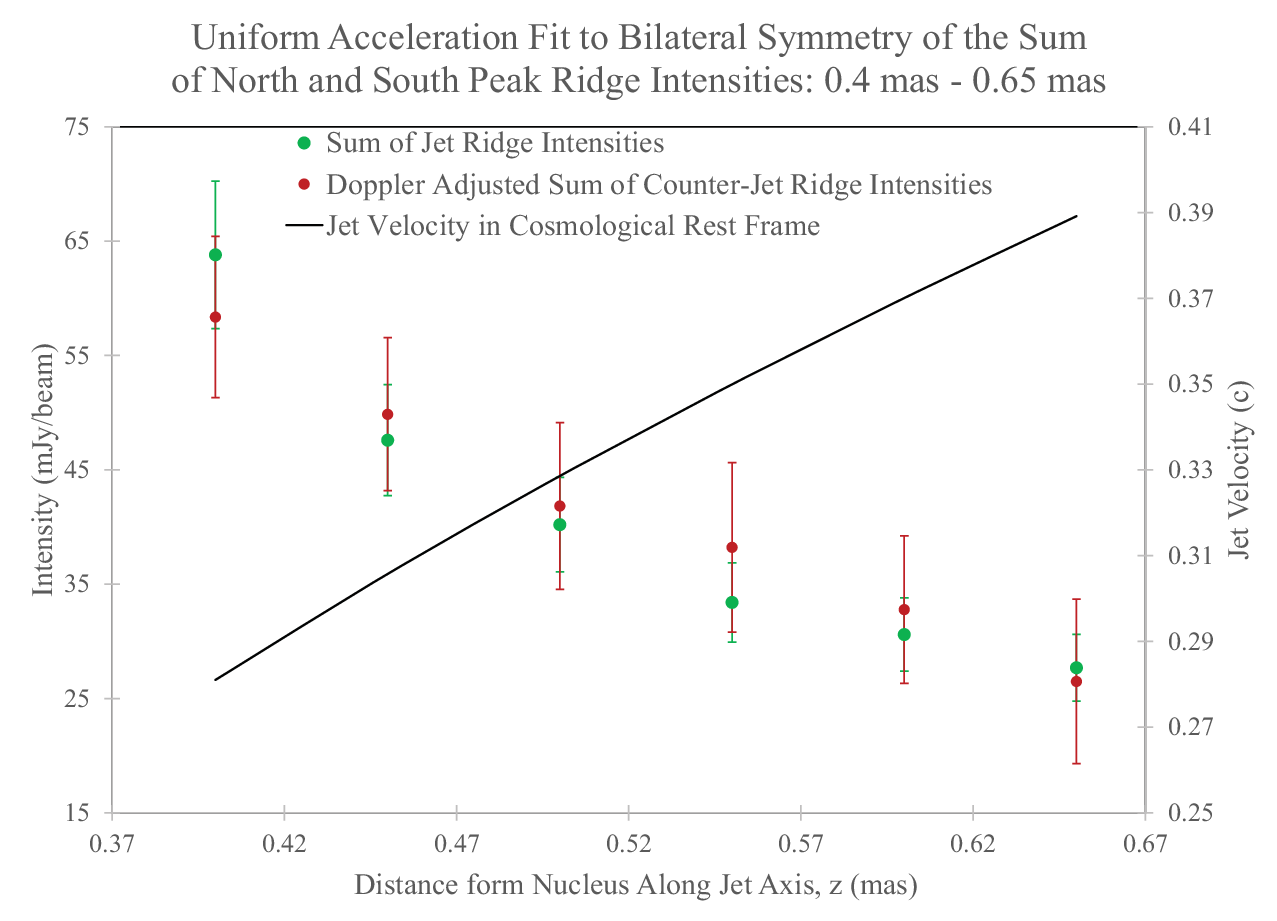}
\includegraphics[width= 0.45\textwidth]{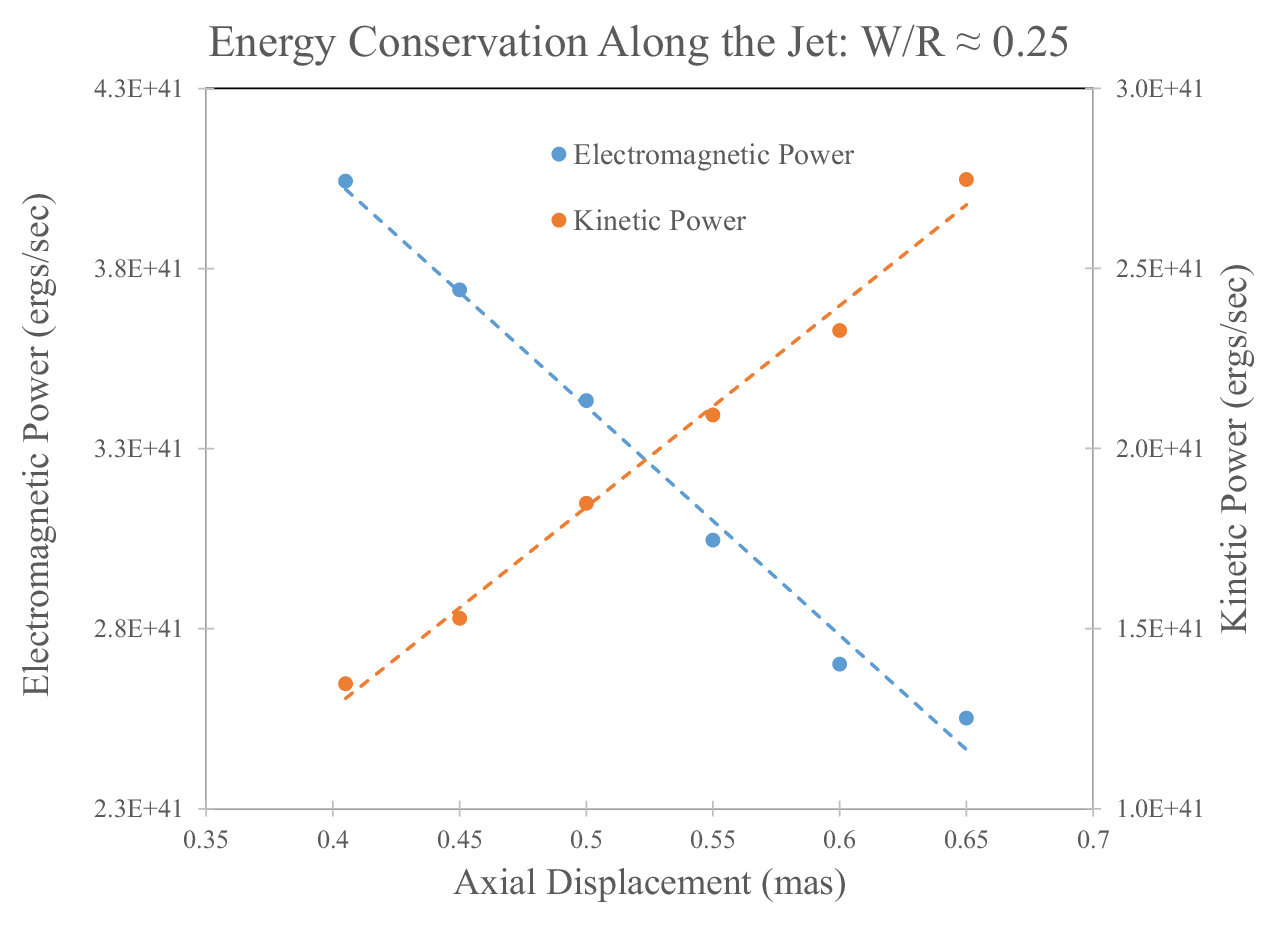}
\includegraphics[width= 0.45\textwidth]{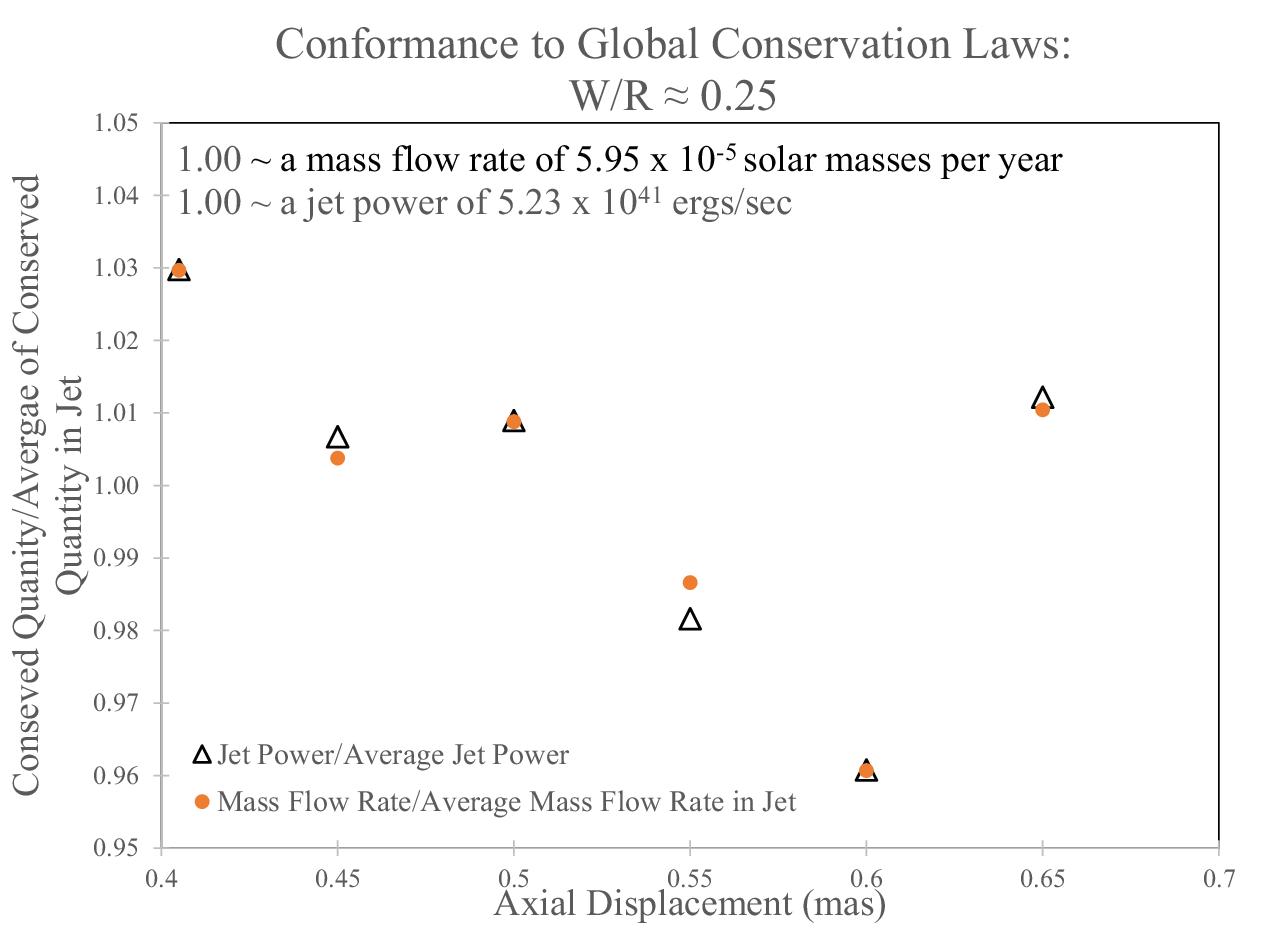}
\includegraphics[width= 0.45\textwidth]{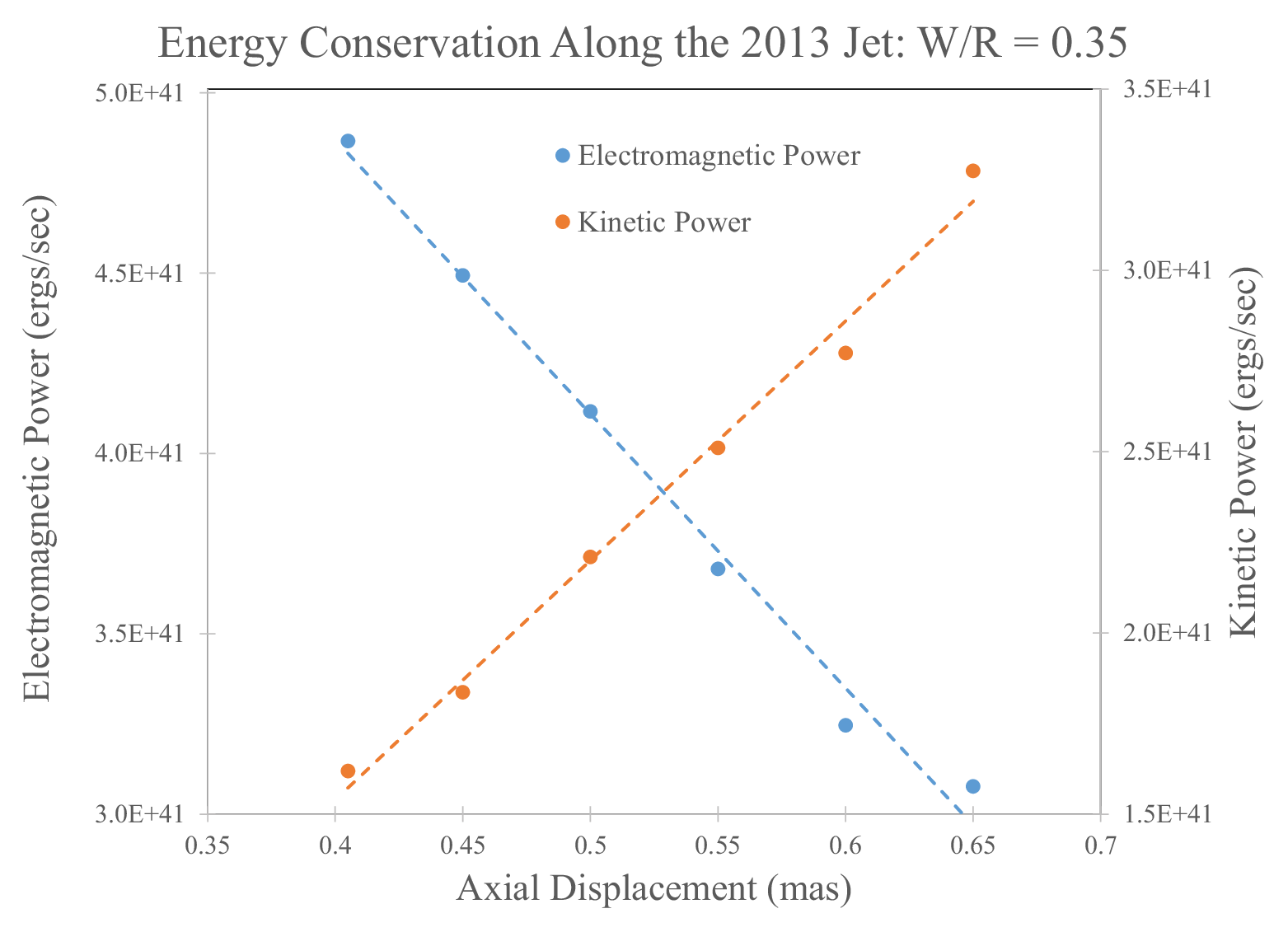}
\includegraphics[width= 0.45\textwidth]{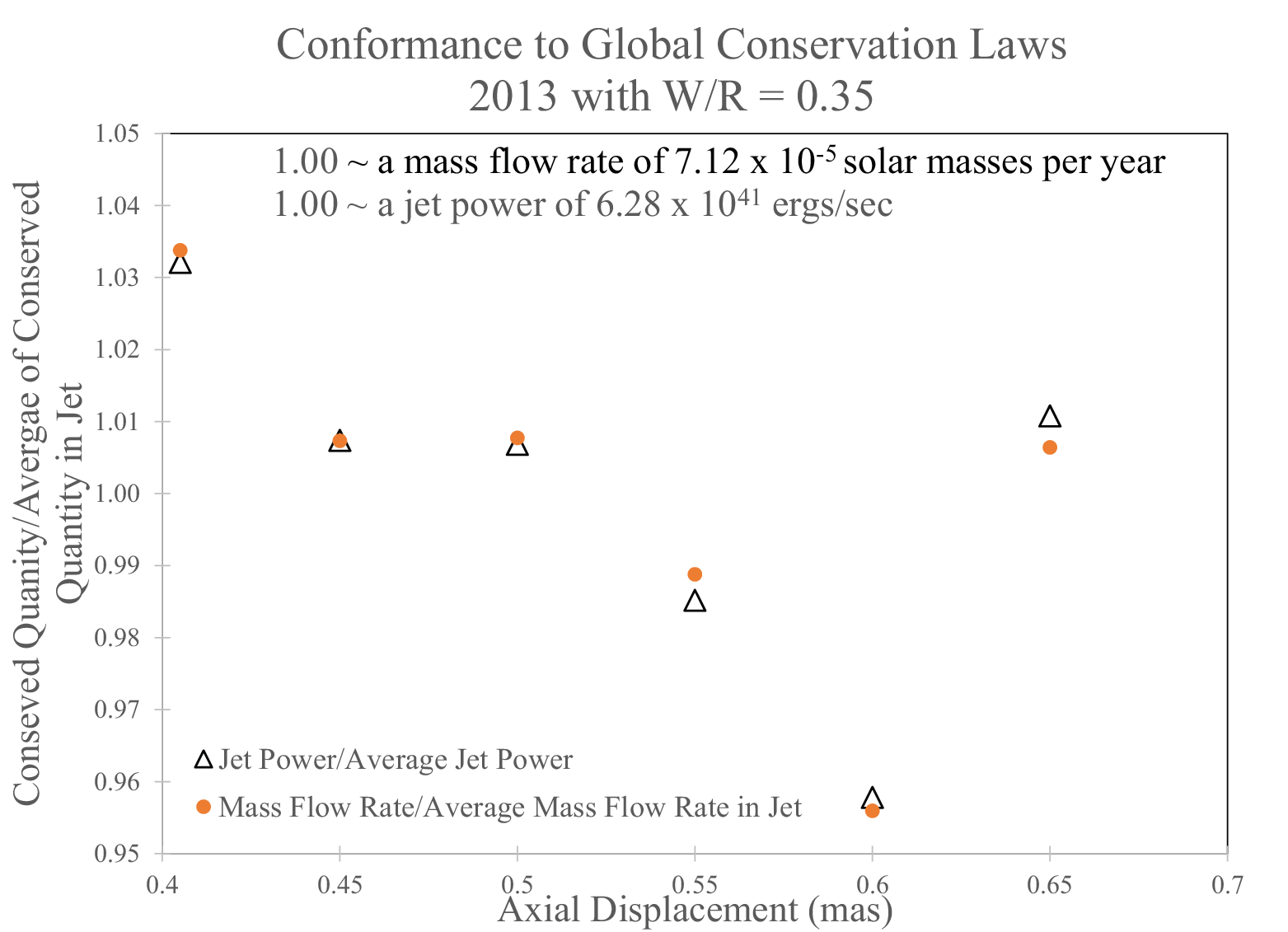}
\caption{Properties of the outer jet in 2013. The top row is reprinted from \citep{pun21}. The left panel is a high-sensitivity 43\,GHz VLBA image from January 12, 2013, that is designed to highlight the counterjet. The white linear contours levels are (1, 1.64, 2.29 ..... 10) mJy/beam. The red contours are at -0.5 mJy/beam. The right panel is the constant acceleration jet velocity profile associated with the Doppler boosts that convert the sum of the ridge peak intensities of the counterjet into those of the approaching jet (see Equation 1). The middle row is reprinted from \citep{pun22}. Poynting flux is converted into kinetic energy along the jet, and the conformance to the conservation laws is tight. Increasing W/R from 0.25 and 0.35 in the bottom row has little effect on the nature of the solution, just the normalization.}
\end{center}
\end{figure*}
\section{Review of relevant equations}
The synchrotron emissivity is given in \citep{tuc75},
\begin{eqnarray}
&& j_{\nu}(\nu_{e}) = 1.7 \times 10^{-21} [4 \pi N_{E}]a(n)B^{(1
+\alpha)}\left(\frac{4
\times 10^{6}}{\nu_{e}}\right)^{\alpha}\;,\nonumber \\
&& \\
&& a(n)=\frac{\left(2^{\frac{n-1}{2}}\sqrt{3}\right)
\Gamma\left(\frac{3n-1}{12}\right)\Gamma\left(\frac{3n+19}{12}\right)
\Gamma\left(\frac{n+5}{4}\right)}
       {8\sqrt\pi(n+1)\Gamma\left(\frac{n+7}{4}\right)} \;,
\end{eqnarray}
where the coefficient $a(n)$ separates the pure dependence on $n$ \citep{gin65}. A power-law energy spectrum for the leptons with normalization $ N_{E}$ and number index $n$ was assumed. The total magnetic field is $B$. A value of $\alpha =0.7$ was chosen for the outer jet because it was previously found that for $z< 1$ mas, $\alpha \approx 0.6$ from 22 GHz to 43 GHz and $\alpha \approx 0.8$ from 43 GHz to 86 GHz \citep{had16,pun21}. This can be transformed into the observed flux density, $S(\nu_{\mathrm{o}})$, for an optically thin jet using the relativistic transformation relations from
\citet{lin85},
\begin{eqnarray}
 && S(\nu_{\mathrm{o}}) = \frac{\delta^{(2 + \alpha)}}{4\pi D^{2}}\int{j_{\nu}^{'}(\nu_{o}) d V{'}}\;,
\end{eqnarray}
where $D$ is the distance. In this expression, the primed frame is the rest frame of the plasma. $\beta$ is the three-velocity of the moving plasma and the Doppler factor, $\delta=1/[\gamma (1-\beta\cos{\theta})]$, with $\theta$ being the LOS, and $\gamma = 1/\sqrt{1-\beta^{2}}$.
\par Consider the time stationary, axisymmetric approximation to the frozen-in conditions in the M\,87 frame \citep{pun08},
\begin{equation}
E^{\perp} = -\beta_{F}B^{P}\;, \; \beta_{F}\equiv (\Omega_{F}R_{\perp})/c\; ,\quad B^{\phi}=\frac{\beta^{\phi}-\beta_{F}}{\beta^{P}}B^{P}\;,
\end{equation}
where P indicates the poloidal direction of $B$, $\perp$ is the orthogonal poloidal direction, $\beta^{i}$ are components of the bulk velocity, and $R_{\perp}$ is the cylindrical radius. The kinematic quantities, $\mathcal{K}(\mathrm{protonic})$, $\mathcal{L}(\mathrm{protonic})$, and $\mathcal{M}$ are the kinetic energy, angular momentum, and mass flux of a protonic plasma, respectively,
\begin{eqnarray}
&& \mathcal{K}(\mathrm{protonic}) = \mathcal{M}c^{2}(\gamma-1)\;, \;\mathcal{L}(\mathrm{protonic}) = \mathcal{M}c\gamma\beta^{\phi}R_{\perp} \;, \nonumber \\
&& \mathcal{M}= Nm_{p}c\gamma \beta^{z}\;,
\end{eqnarray}
where $m_{p}$ is the mass of the proton, and $N$ is the proper number density. The lepto-magnetic energy, $E(\mathrm{lm})$, is the volume integral of the leptonic internal energy density, $U_{e}$, and the magnetic field energy density, $U_{B}$,
\begin{eqnarray}
 && E(\mathrm{lm}) = \int{(U_{B}+ U_{e})}\, dV =  \nonumber \\
&& \int{\left[\frac{B^{2}}{8\pi}+ \int_{E_{min}}^{E_{max}}(m_{e}c^{2})(N_{E}E^{-n + 1})\, d\,E \right]dV}, \,
\end{eqnarray}
where $dV$ is the volume element. The leptonic contributions are negligible to the solutions relevant to this study. For consistency with the outer jet solutions, $E_{min} \approx m_{e}c^{2}$ in the cylinder \citep{pun22}. $E_{max}$ has very little effect on the solutions.
\par The electromagnetic angular momentum, $S_{L}$, and the poloidal Poynting power, $S^{P}$, are \citep{pun08}
\begin{eqnarray}
&& S_{L} = \frac{1}{4\pi}\int{B^{T}B^{P}dA_{\perp}}\;, S^{P} = \frac{1}{4\pi}\int{\Omega_{F}B^{T}B^{P}dA_{\perp}}\;,\nonumber \\
&& \Phi \equiv \int{B^{P}dA_{\perp}}\;,
\end{eqnarray}
where
\begin{eqnarray}
&& B^{T}=[\sqrt{r^{2}-2Mr+a^{2}}\sin{\theta}]\sqrt{F^{r\,\theta}F_{r\, \theta}} \;, \\ \nonumber
&& \;-M\leq a \leq M\;,
\end{eqnarray}
where $F^{\mu\,\nu}$ is the Faraday tensor. Combining Equations (7) and (9) in the M87 frame,
\begin{eqnarray}
&& Q\approx\int{\mathcal{K}(\mathrm{protonic})dA_{\perp}}+S^{P}\;,  \nonumber \\
&& L\approx\int{\mathcal{L}(\mathrm{protonic})dA_{\perp}}+S_{L}\, \quad \dot{M}=\int{\mathcal{M}(\mathrm{protonic})dA_{\perp}}\;. \nonumber \\
&&
\end{eqnarray}
\section{Conformance plots} This appendix provides conservation law conformance plots analogous to those used to analyze solution viability in \citet{pun22} and Appendix A.
\begin{figure*}
\begin{center}
\includegraphics[width= 0.45\textwidth]{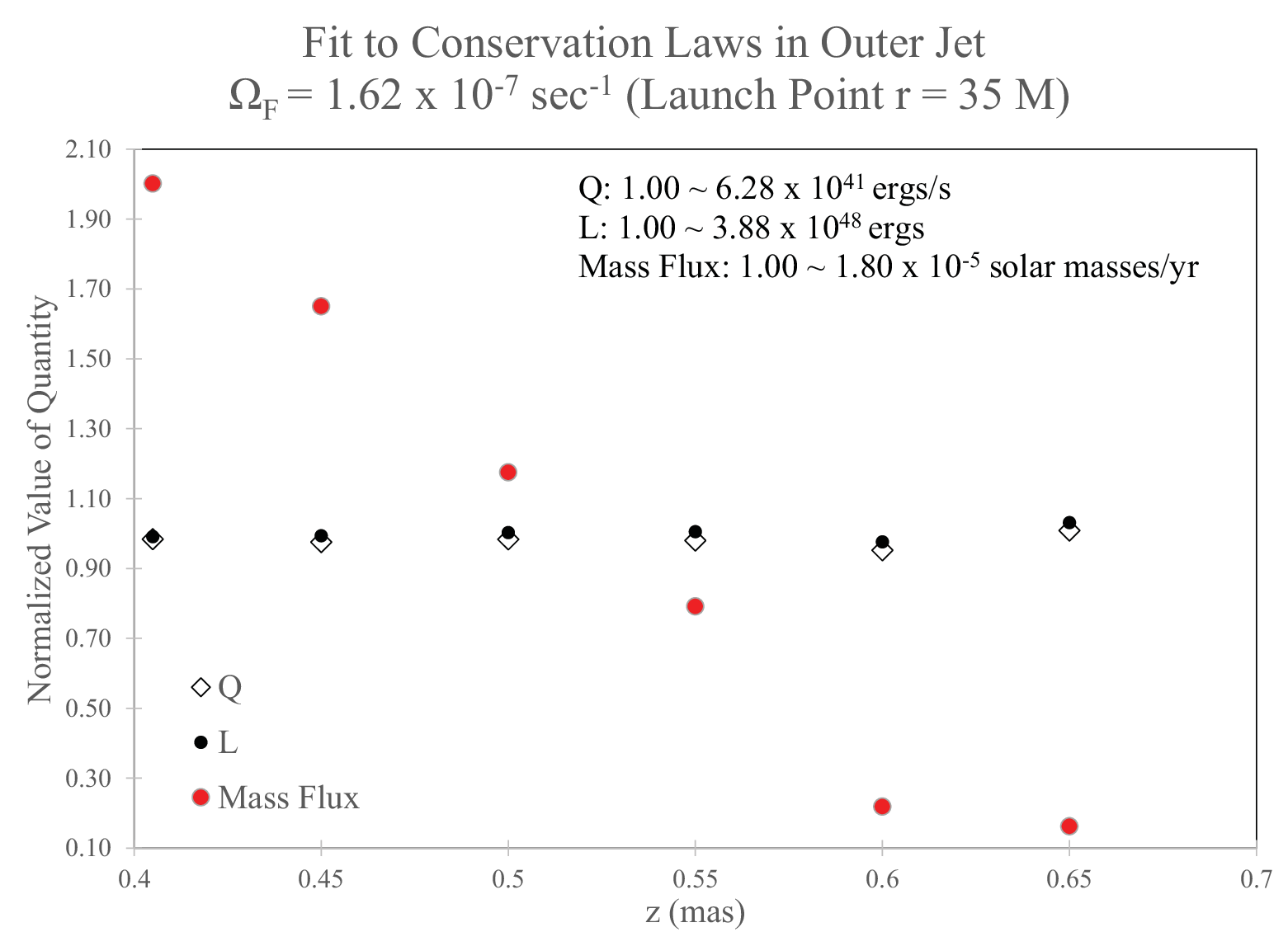}
\includegraphics[width= 0.45\textwidth]{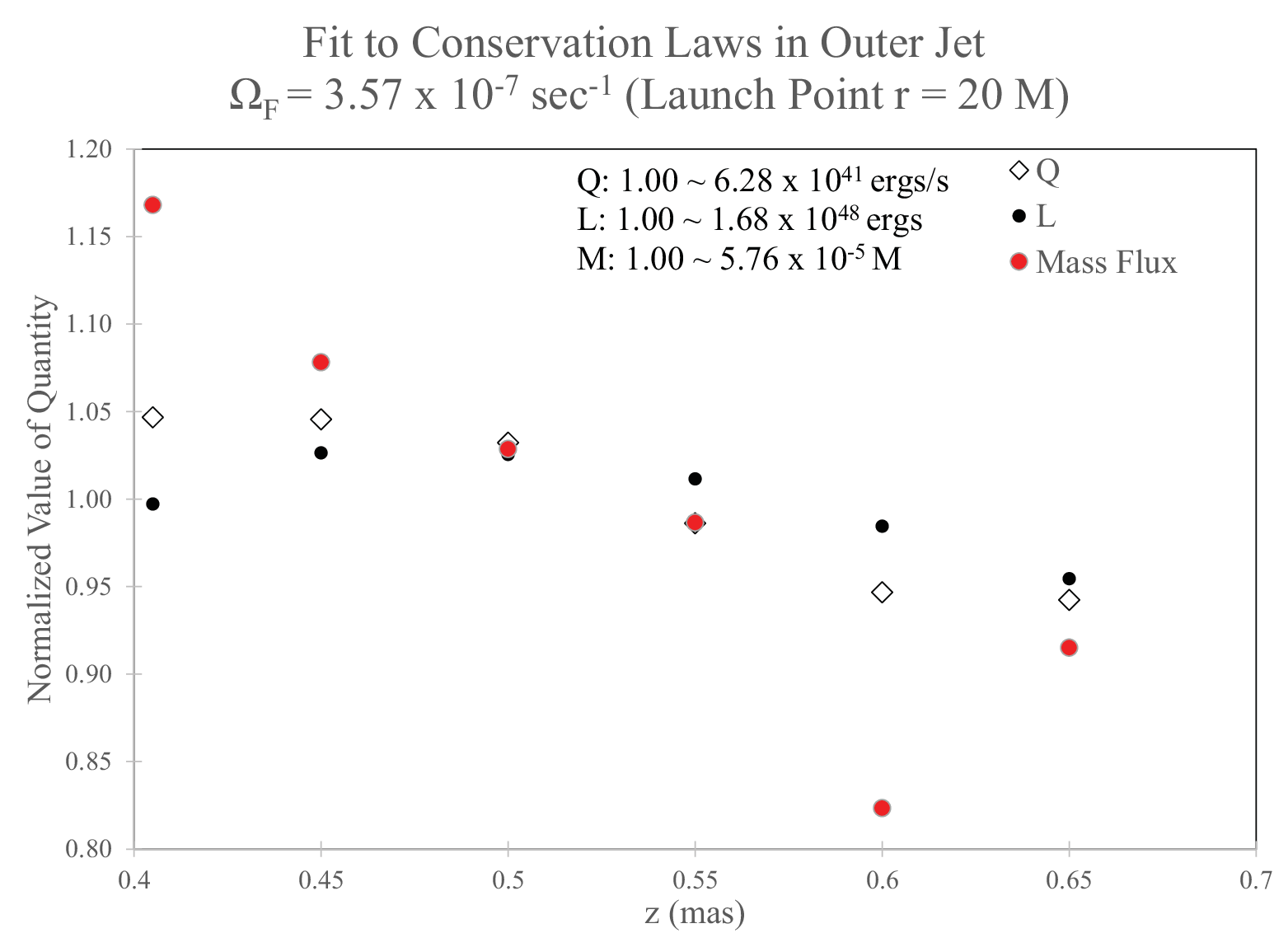}
\includegraphics[width= 0.45\textwidth]{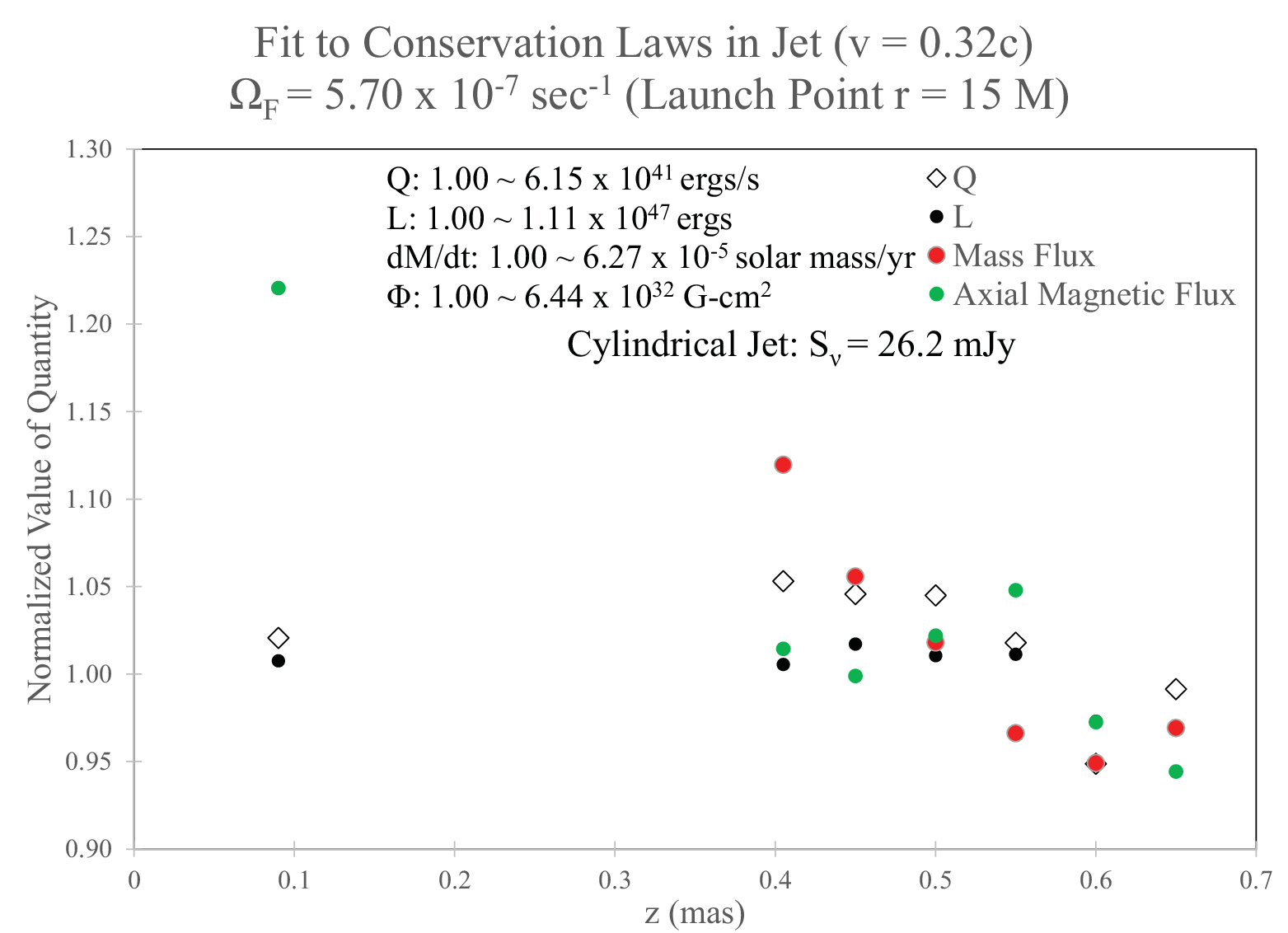}
\includegraphics[width= 0.45\textwidth]{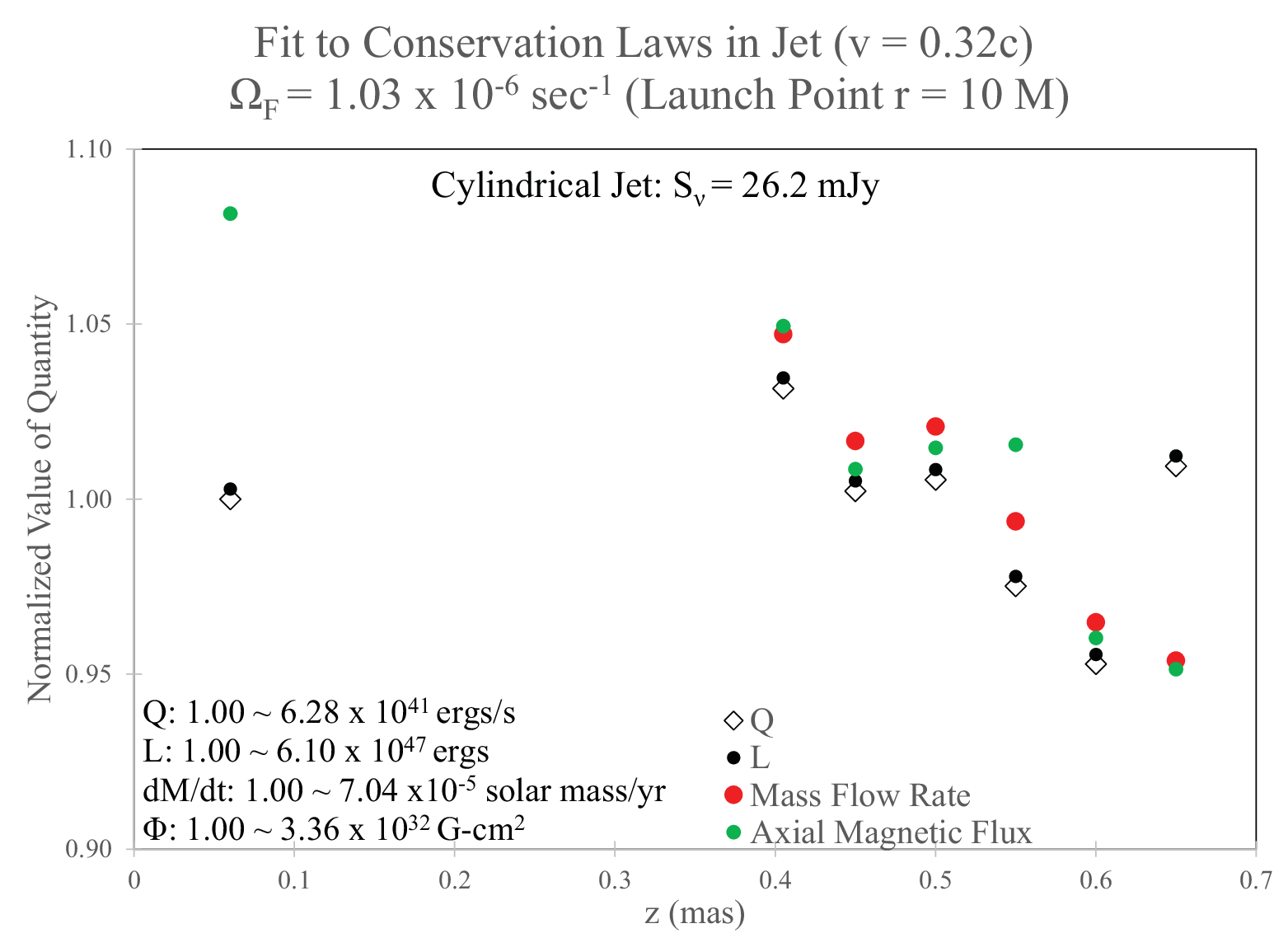}
\includegraphics[width= 0.45\textwidth]{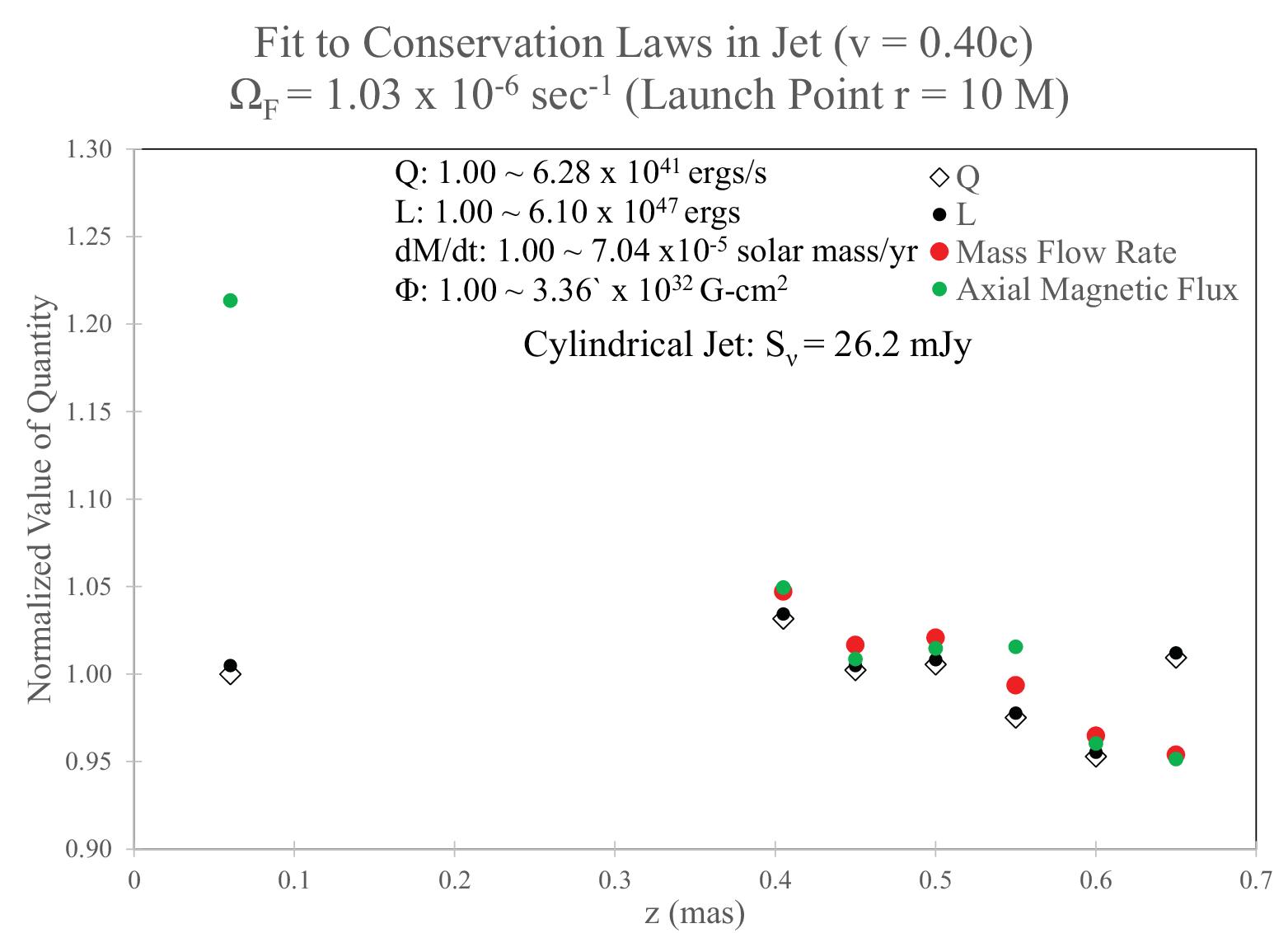}
\caption{Finding a viable solution. The four panels in the first two rows show that mass flux conservation in the outer jet improves as the launch point moves inward. At r=10M, the mass flux is conserved in the outer jet and the magnetic flux is marginally conserved globally for v=0.32c, but not for v=0.4c in the bottom panel. }
\end{center}
\end{figure*}
\begin{figure*}
\begin{center}
\includegraphics[width= 0.45\textwidth]{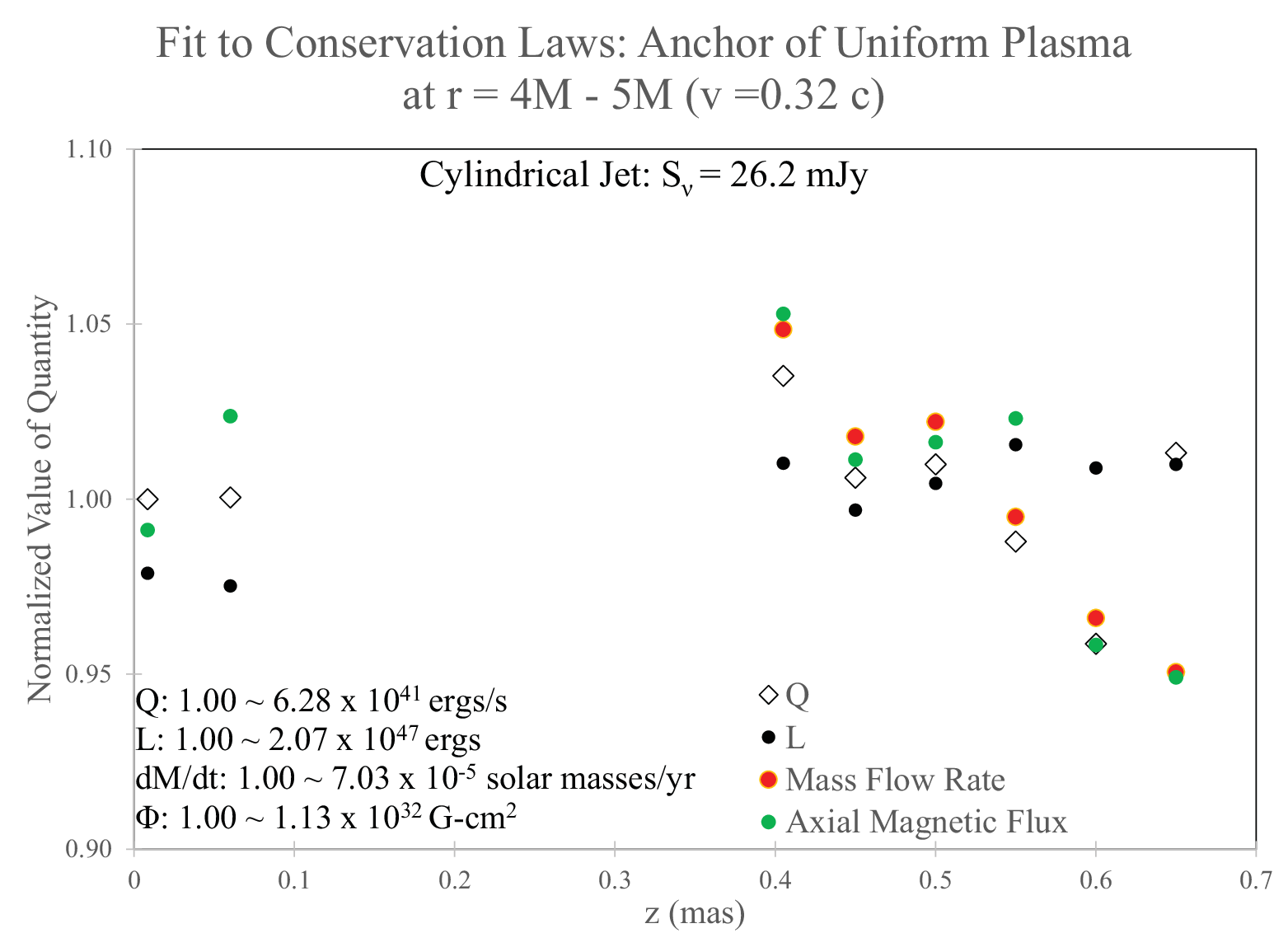}
\includegraphics[width= 0.45\textwidth]{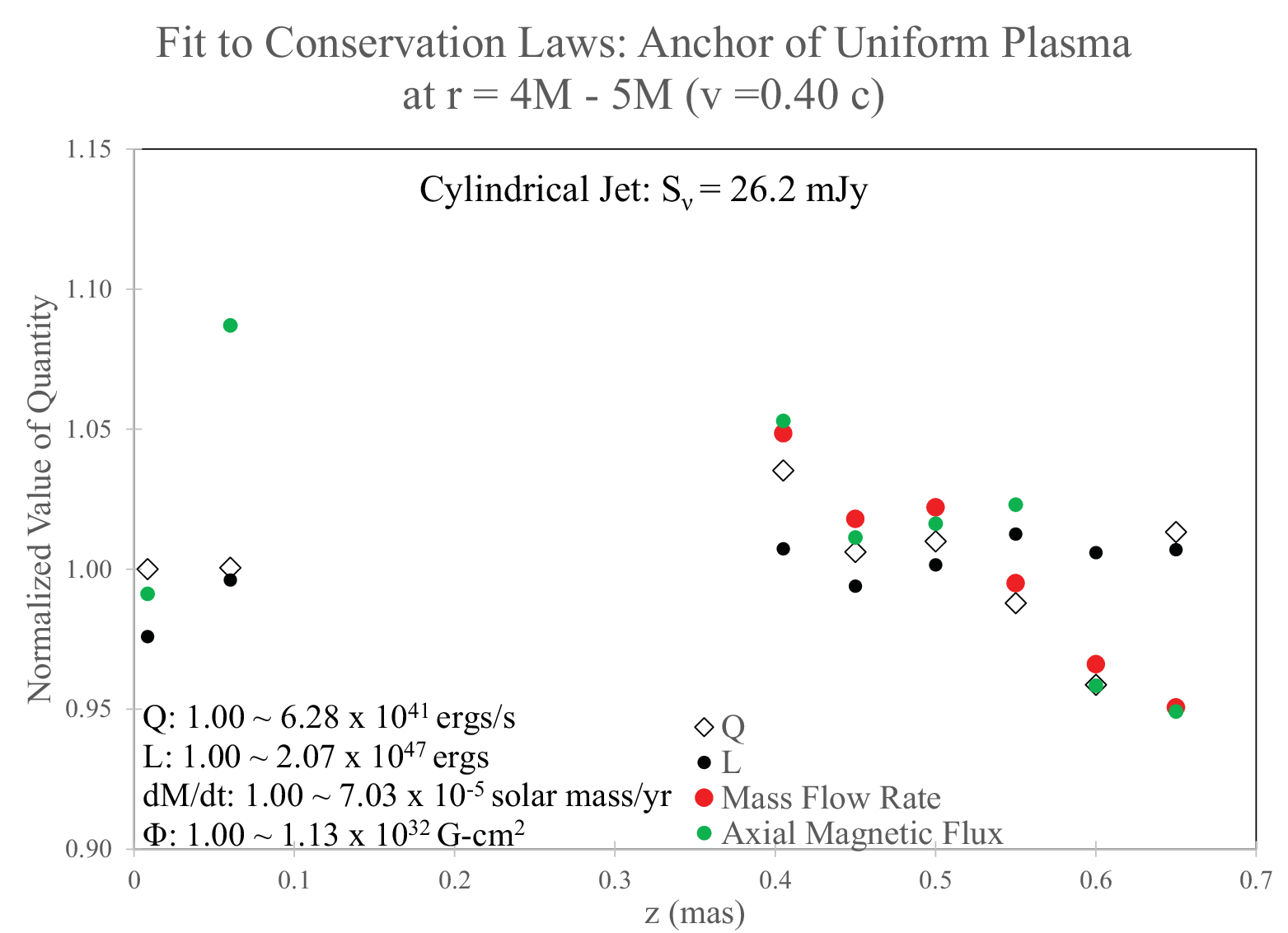}
\includegraphics[width= 0.45\textwidth]{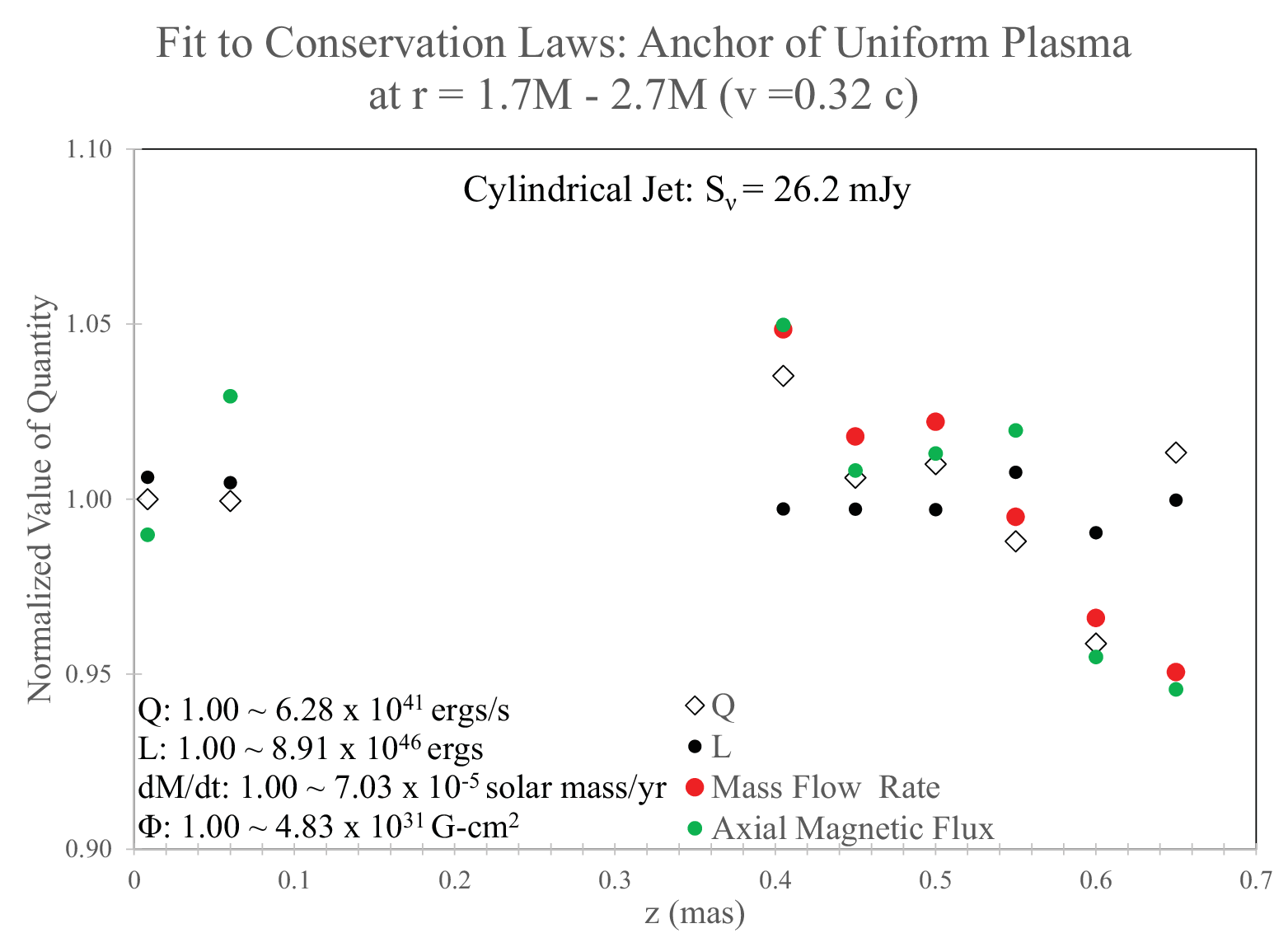}
\includegraphics[width= 0.45\textwidth]{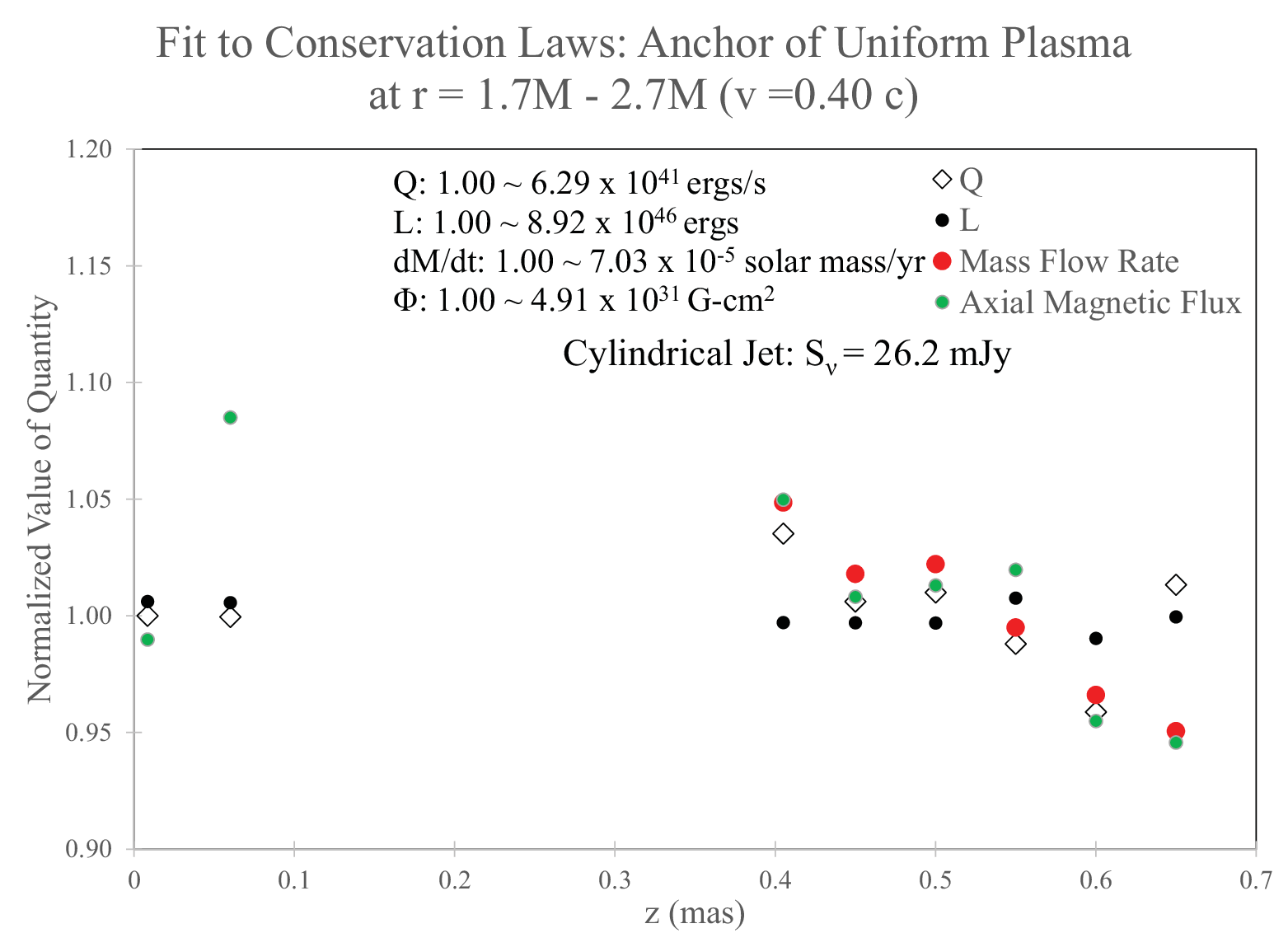}
\caption{Moving the launch point inward. The first row shows improvement by an annulus of plasma between 4M and 5M in the $\Phi$ conservation compared to $r=10M$ in Figure C.1. The bottom row shows no noticeable change in conformance at smaller radii. }
\end{center}
\end{figure*}
\section{Uncertainty analysis of the flux density}
In this appendix, the variation in the conformance plots as a function of the cylindrical jet flux density is explored, $26.2^{+8.8}_{-7.1}$ mJy.

\begin{figure*}
\begin{center}
\includegraphics[width= 0.45\textwidth]{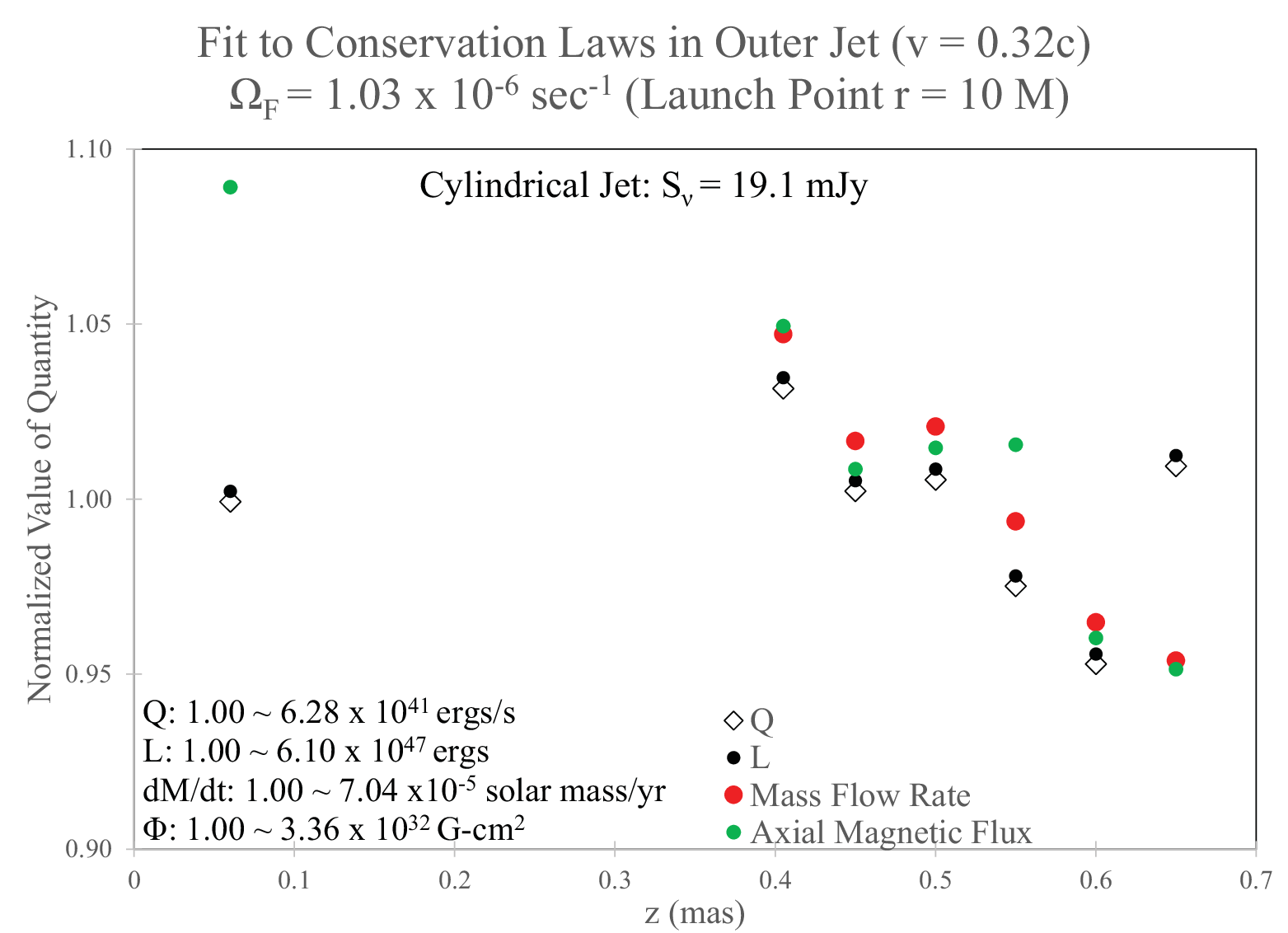}
\includegraphics[width= 0.45\textwidth]{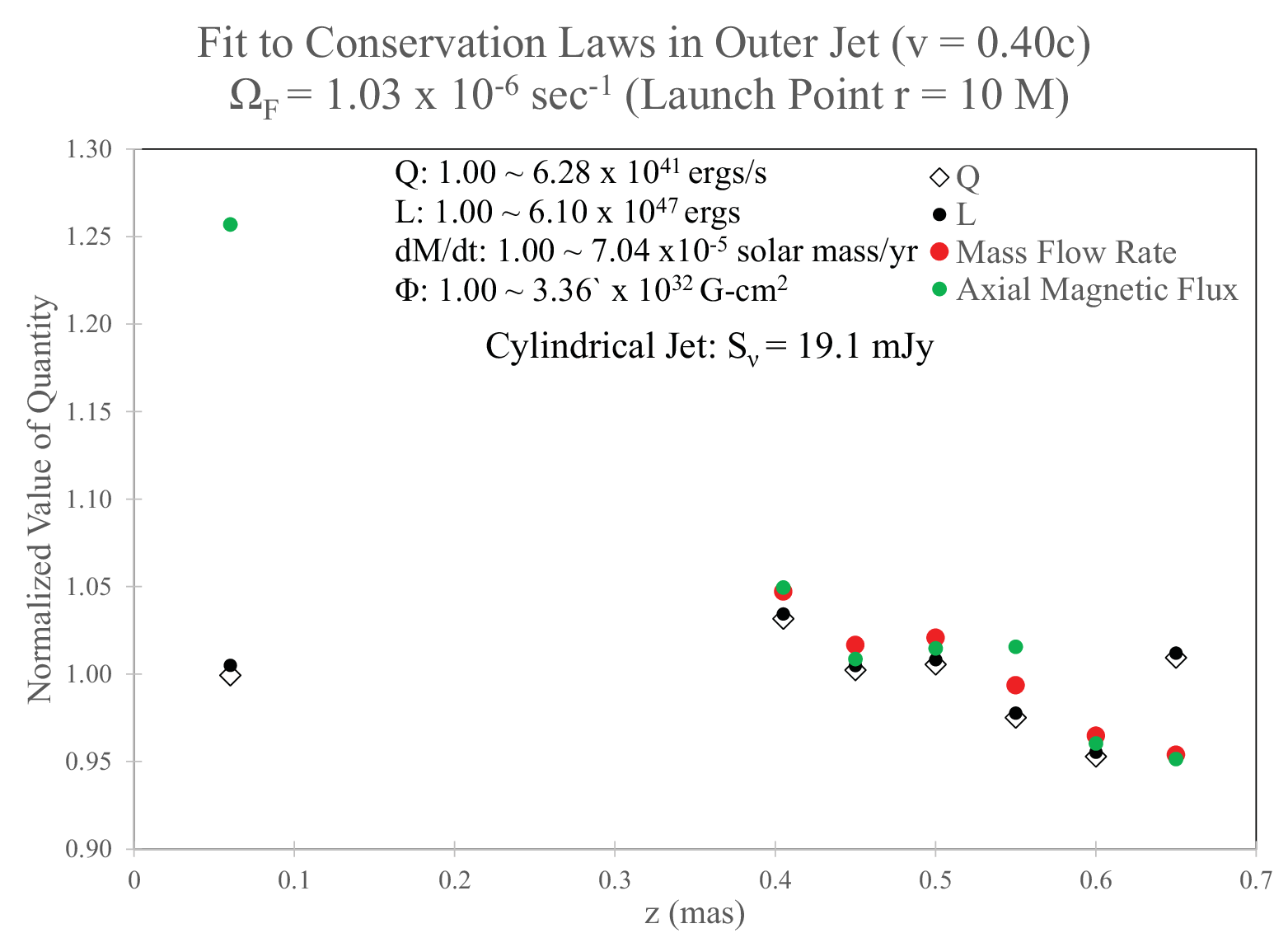}
\includegraphics[width= 0.45\textwidth]{f20.eps}
\includegraphics[width= 0.45\textwidth]{f21.eps}
\includegraphics[width= 0.45\textwidth]{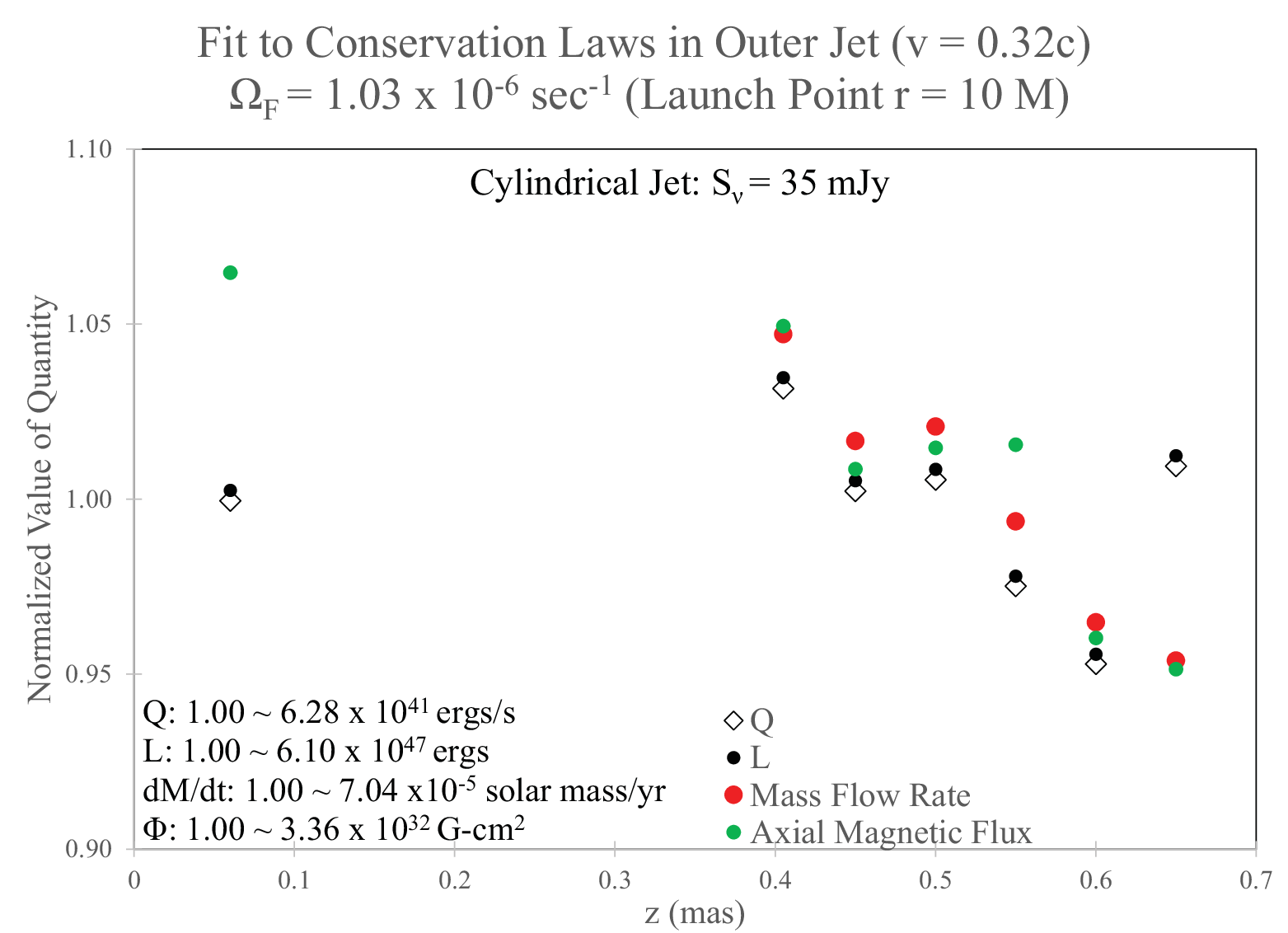}
\includegraphics[width= 0.45\textwidth]{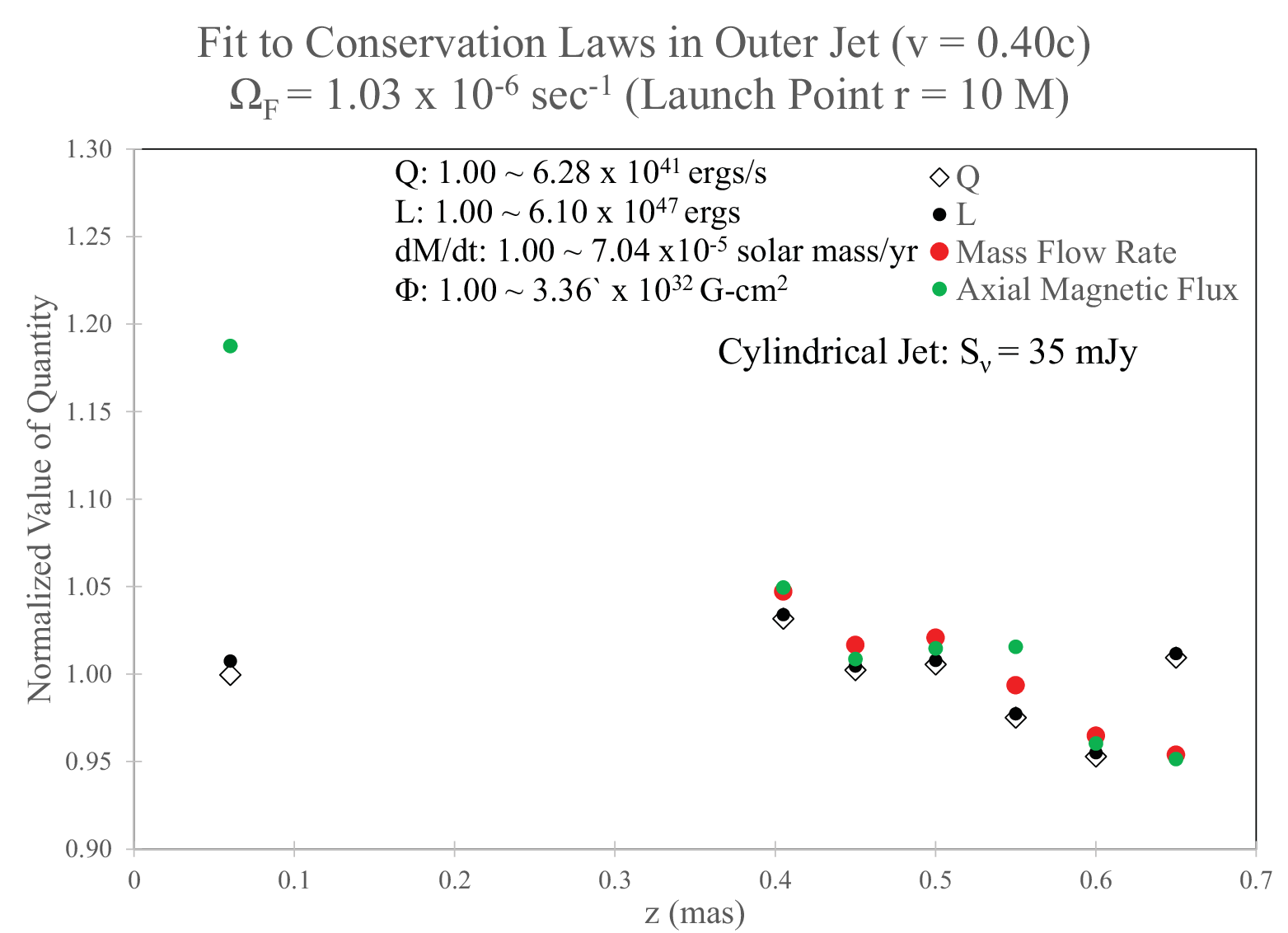}
\caption{Flux density variation for an anchor at r =10M. The fit to $\Phi$ conformance is improved with a higher flux density. r = 10 M conforms marginally even at 35 mJy and $v =0.32$c, however. The identification of r = 10M as the outermost anchor point that is viable is not a consequence of the flux density uncertainty.}
\end{center}
\end{figure*}
\begin{figure*}
\begin{center}
\includegraphics[width= 0.45\textwidth]{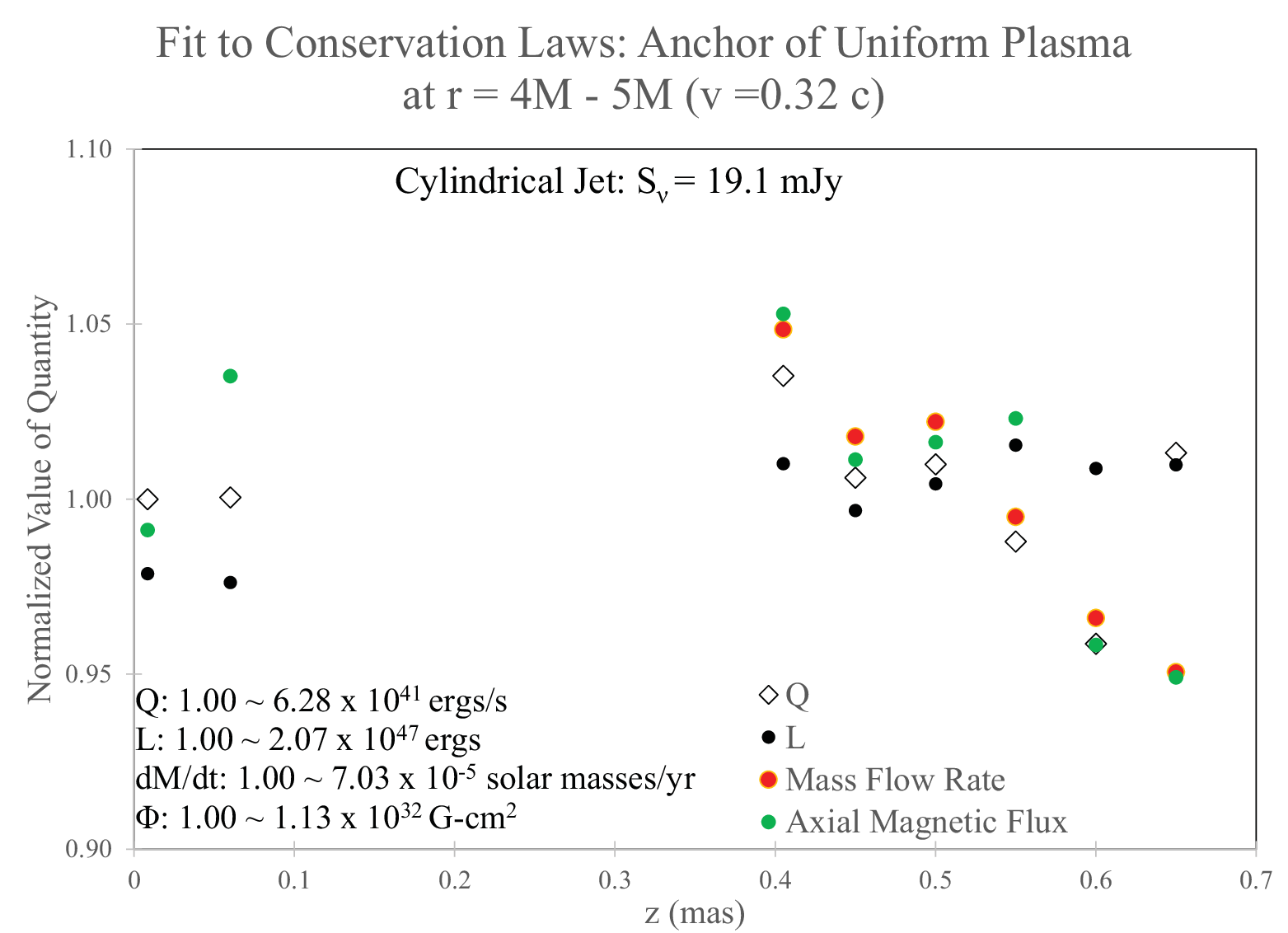}
\includegraphics[width= 0.45\textwidth]{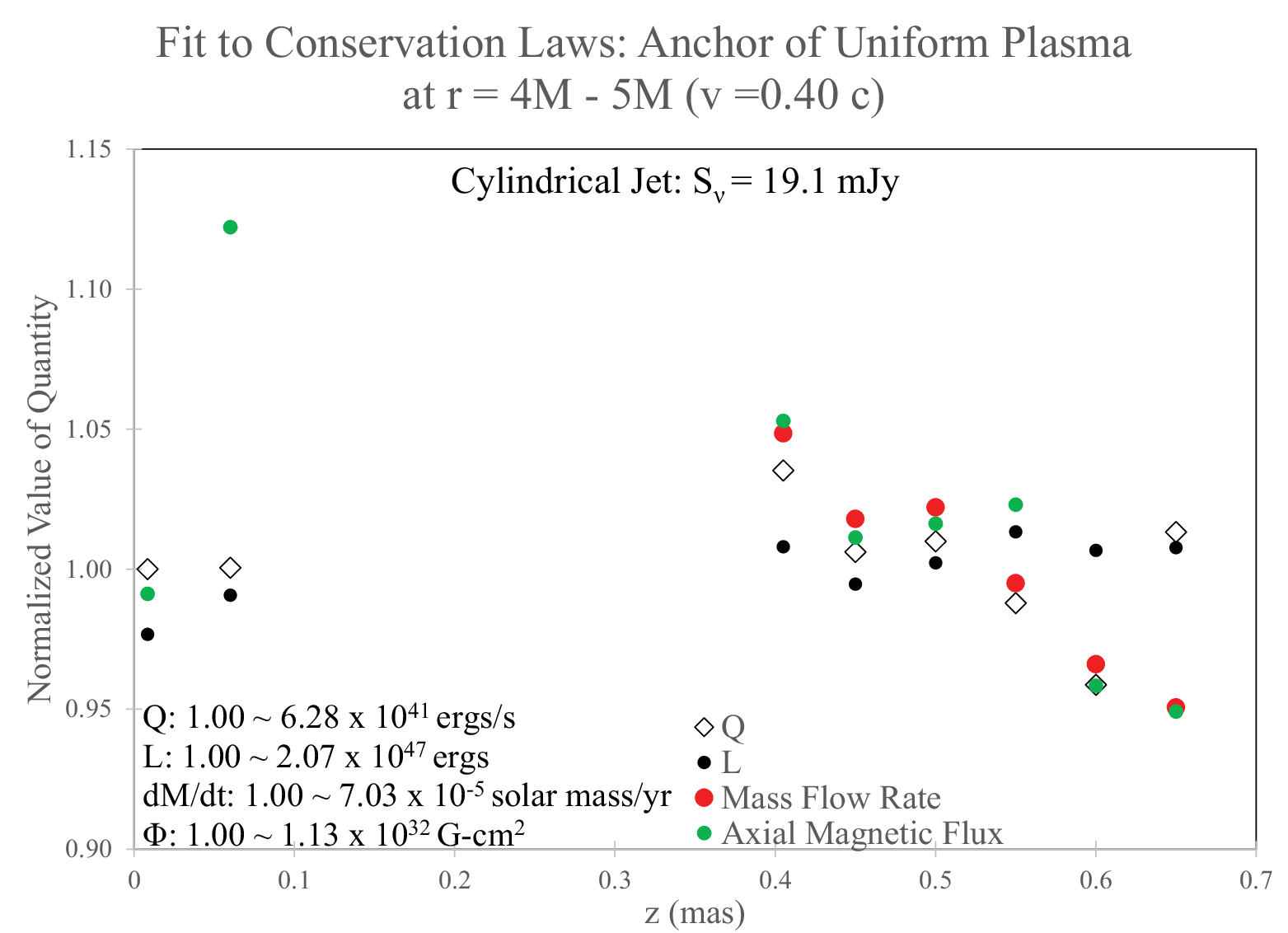}
\includegraphics[width= 0.45\textwidth]{f22.eps}
\includegraphics[width= 0.45\textwidth]{f23.eps}.
\includegraphics[width= 0.45\textwidth]{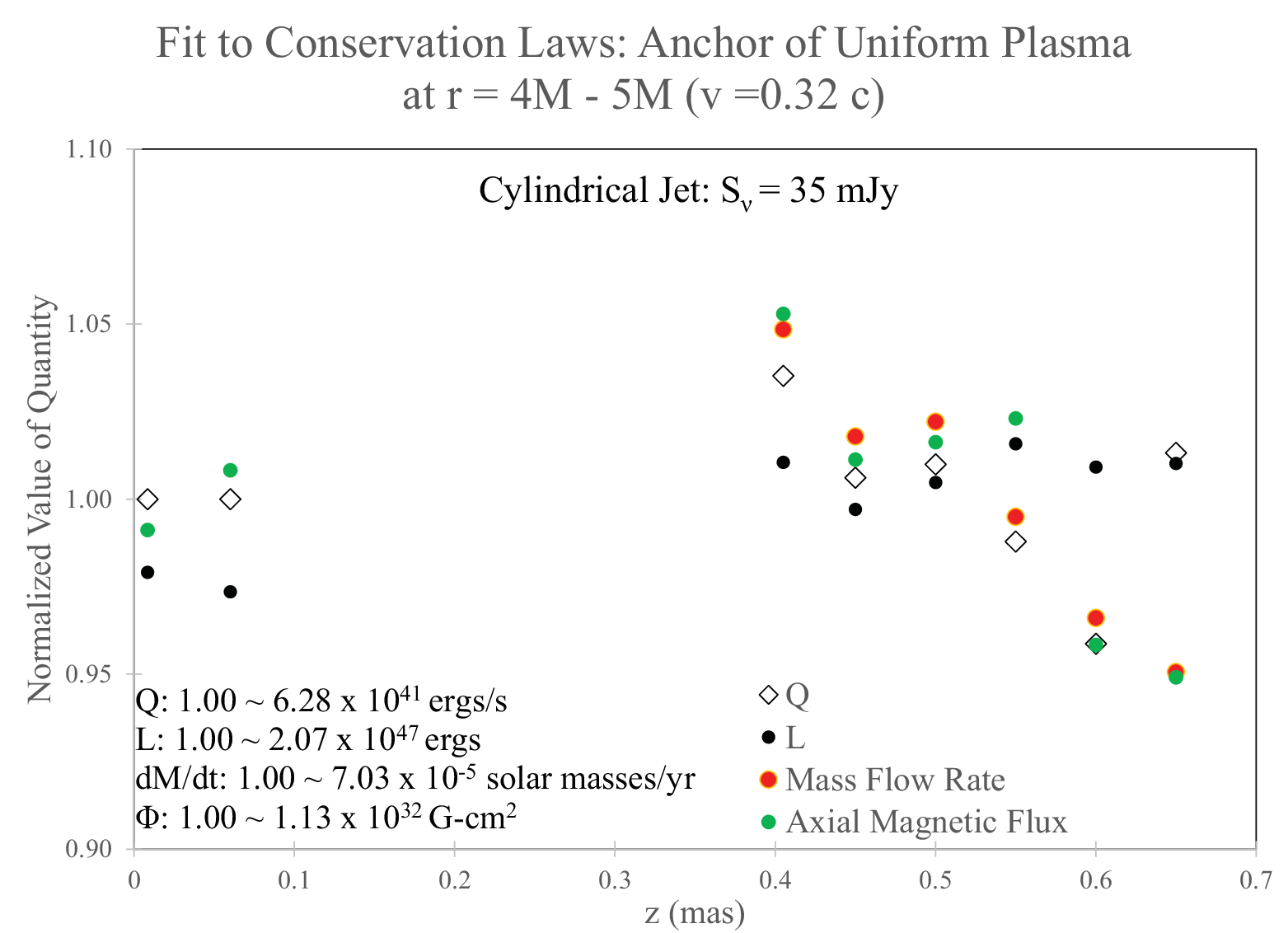}
\includegraphics[width= 0.45\textwidth]{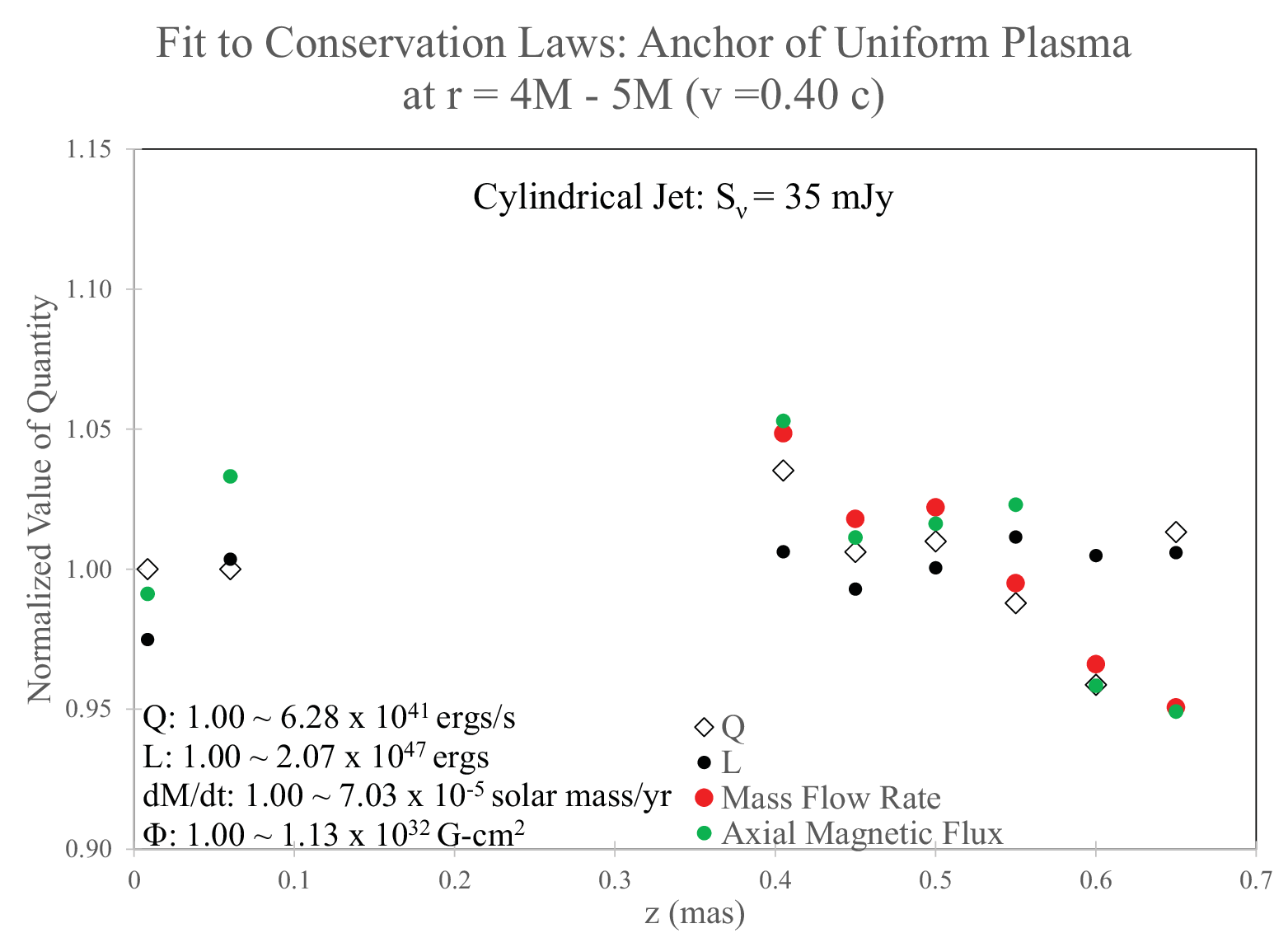}
\caption{Flux density variation for an annular anchor at r =4M-5M. The fit to $\Phi$ conformance is improved with a higher flux density. An actual model of the anchor, the annular region, is required to capture gradients in the Kerr geometry as the event horizon is approached. }
\end{center}
\end{figure*}
\begin{figure*}
\begin{center}
\includegraphics[width= 0.45\textwidth]{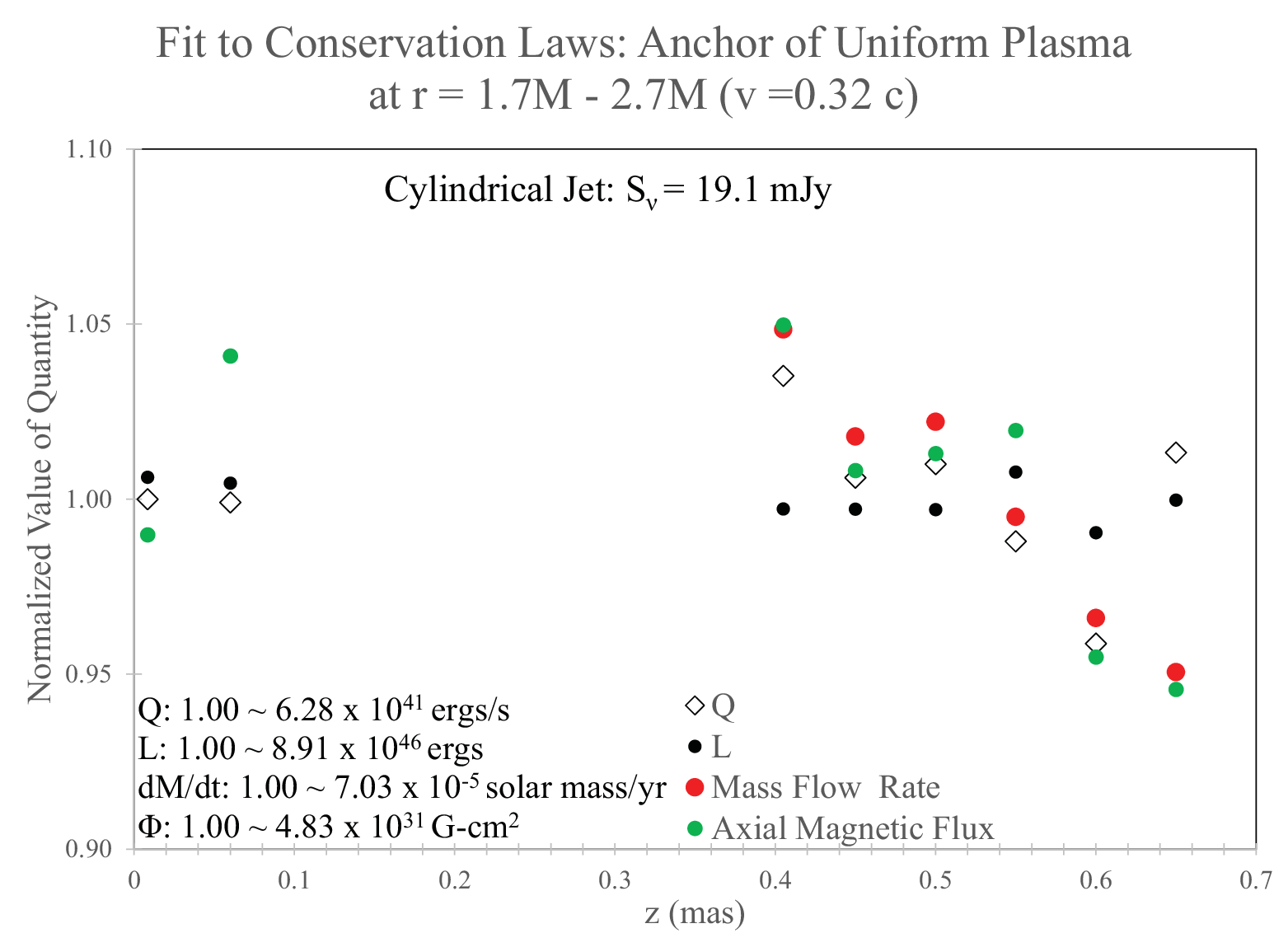}
\includegraphics[width= 0.45\textwidth]{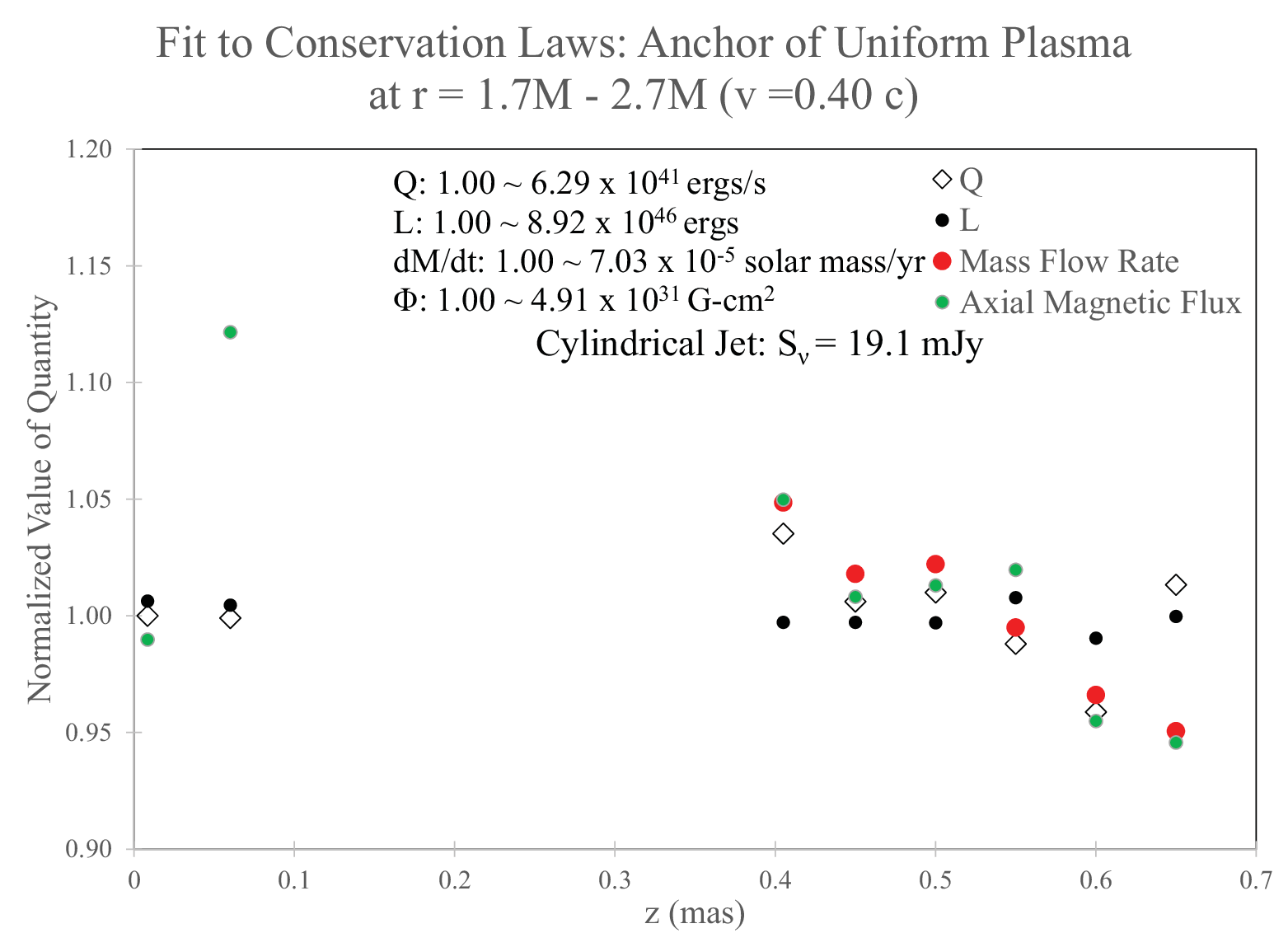}
\includegraphics[width= 0.45\textwidth]{f24.eps}
\includegraphics[width= 0.45\textwidth]{f25.eps}
\includegraphics[width= 0.45\textwidth]{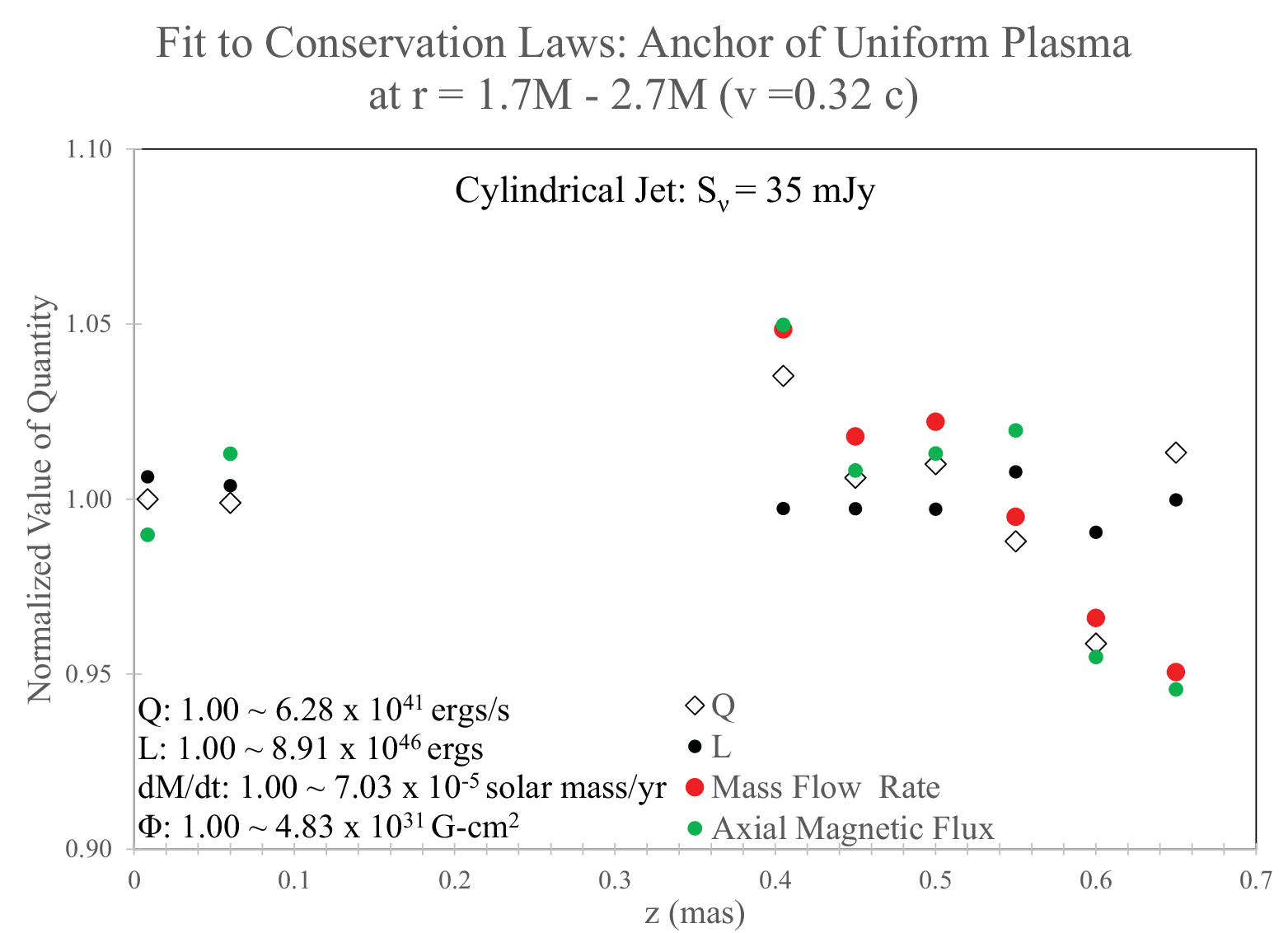}
\includegraphics[width= 0.45\textwidth]{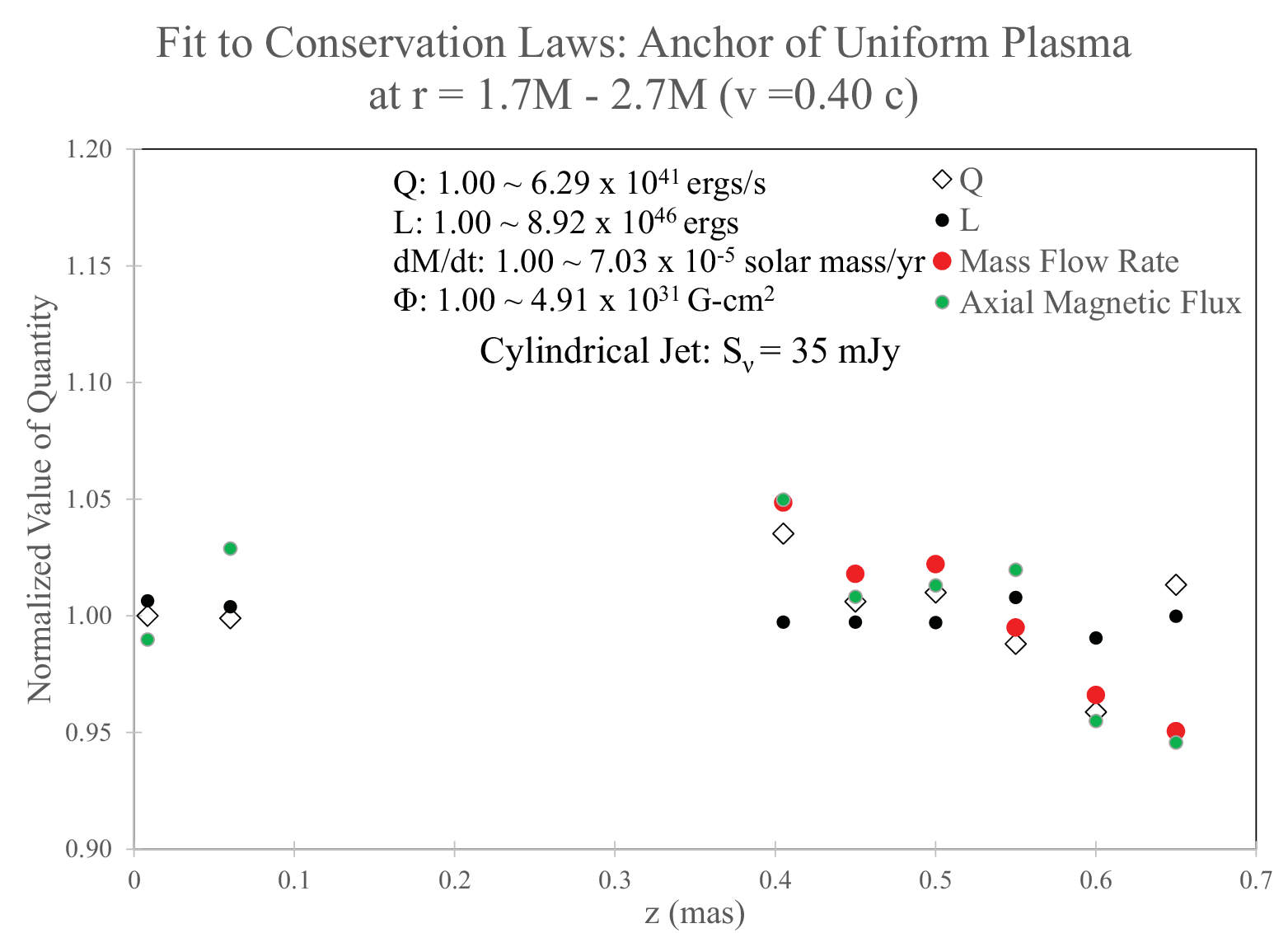}
\caption{Flux density variation for an annular anchor at r =1.7M-2.7M. The fit to $\Phi$ conformance is improved with a higher flux density. }
\end{center}
\end{figure*}

\end{appendix}

\begin{thebibliography}{}
\bibitem[Algaba et al.(2021)]{alg21}Algaba,, J., Anczarski, J. Asada, K. et al. 2021, ApJL, 911, L11
\bibitem[Bicknell(1995)]{bic95}Bicknell, G. 1995, ApJSS, 101, 29
\bibitem[Blandford and Payne(1982)]{bla82}Blandford, R. D., \& Payne, D. G. 1982, MNRAS, 199, 883
\bibitem[EHT Collaboration et al.(2019)]{eht19} Event Horizon Telescope Collaboration, Akiyama, K., Alberdi, A., et al. 2019, ApJL, 875, L5
\bibitem[EHT Collaboration et al.(2021)]{aki22} Event Horizon Telescope Collaboration, Akiyama, K., Algaba, J. C., et al. 2021, ApJL, 910, L13
\bibitem[Fomalont(1999)]{fom99} Fomalont, E. 1999, in Synthesis Imaging in Radio Astronomy II, ASP Conference
Series, eds. Taylor, G., Carilli, C., Perley, R. 180, 301
\bibitem[Ginzburg and Syrovatskii(1965)]{gin65} Ginzburg, V. and Syrovatskii, S. 1965,
  Annu. Rev. Astron. Astrophys. 3 297
\bibitem[Hada et al.(2013)]{had12} Hada, K., Kino, M., Niinuma, K., et al. 2013, PASJ, 65, 24
\bibitem[Hada et al.(2014)]{had14} Hada, K., Giroletti, M., Kino, M., et al. 2014, ApJ, 788, 165
\bibitem[Hada et al.(2016)]{had16} Hada, K., Kino, M., Doi, A., et al. 2016, ApJ, 817, 131
\bibitem[Johnson et al.(2023)]{joh23}Johnson, M.., Akiyama, K., Blackburn, L. et al 2023 Galaxies, 11(3), 61
\bibitem[Kim et al.(2018)]{kim18} Kim, J.-Y., Lee, S.-S.., Hogsdon, J., et al. 2018, A\&A, 610, L5
\bibitem[Lee et al.(2008)]{lee08} Lee, S.-S., Lobanov, A., Krichbaum, T. P. et al. 2008, ApJ, 136, 59
\bibitem[Lightman et al.(1975)]{lig75}Lightman, A., Press, W., Price, R. and Teukolsky, S. 1975, \emph{Problem Book in Relativity and Gravitation} (Princeton University Press, Princeton)
\bibitem[Lind and Blandford(1985)]{lin85}Lind, K., Blandford, R.1985 ApJ 295 358
\bibitem[Lu et al.(2023)]{lu23} Lu, RS., Asada, K., Krichbaum, T.P. et al. Nature 616, 686–690 (2023). https://doi.org/10.1038/s41586-023-05843-w
\bibitem[Punsly(2008)]{pun08} Punsly, B. 2008, \emph{Black Hole Gravitohydromagnetics}, second edition (Springer-Verlag, New York)
\bibitem[Punsly(2021)]{pun21} Punsly, B. 2021, ApJ 918, 4
\bibitem[Punsly(2022)]{pun23}Punsly, B. 2022, ApJ 936, 79
\bibitem[Punsly(2023)]{pun24}Punsly, B. 2023, A\&A 679, L1
\bibitem[Punsly and Chen(2021)]{pun22}Punsly, B. and Chen, S. 2021, ApJL 921 L38
\bibitem[Rohoza et al.(2024)]{roh23} Rohoza, V., Lalakos, A., Paik, M. et al. 2024, arXiv231100018R
\bibitem[Ripperda et al.(2022)]{rip22} Ripperda, B., Liska, M., Chatterjee, K. et al. 2022, ApJ, 924, 32
\bibitem[Tucker(1975)]{tuc75}Tucker, W. 1975, \emph{Radiation Processes in Astrophysics} (MIT Press, Cambridge).
\bibitem[Walker et al.(2018)]{wal18} Walker R. C., Hardee P., Davies, F., Ly, C., Junor, W., 2018, ApJ 855 12\emph{}
\end{thebibliography}
\end{document}